\documentclass[11pt,nofootinbib,notitlepage,superscriptaddress]{revtex4-1}
\usepackage{amsmath,amssymb,float,color}
\usepackage[pdftex]{graphicx}

\def\bq{{\boldsymbol q}}
\def\bp{{\boldsymbol p}}
\def\bE{{\boldsymbol E}}
\def\bB{{\boldsymbol B}}

\def\pr{\prime}
\def\cd{\!\cdot\!}
\def\bxp{{\boldsymbol x}_{\!\perp}}
\def\bpp{{\boldsymbol p}_{\!\perp}}
\def\bqp{{\boldsymbol q}_{\!\perp}}

\def\bgam{{\boldsymbol \gamma}}
\def\ap{a_{\!\perp}}
\newcommand{\ltsim}{\protect\raisebox{-0.5ex}{$\:\stackrel{\textstyle <}{\sim}\:$}}
\newcommand{\gtsim}{\protect\raisebox{-0.5ex}{$\:\stackrel{\textstyle >}{\sim}\:$}}

\begin{document} 
\count\footins = 1000

\title{Nonequilibrium axial charge production in expanding glasma flux tubes}

\author{Naoto Tanji}\email{ntanji@ectstar.eu}
\affiliation{European Centre for Theoretical Studies in Nuclear Physics and Related Areas (ECT*) and Fondazione Bruno Kessler, Strada delle Tabarelle 286, I-38123 Villazzano (TN), Italy}

\date{\today}

\begin{abstract}
Axial charge production at the early stage of heavy-ion collisions is investigated within the framework of real-time lattice simulations at leading order in QCD coupling.
Starting from color glass condensate initial conditions, the time evolution of quantum quark fields under classical color gauge fields is computed on a lattice in longitudinally expanding geometry. 
We consider simple color charge distributions in Lorentz contracted nuclei that realize flux tube-like configurations of color fields carrying nonzero topological charge after a collision. 
By employing the Wilson fermion extended to the longitudinally expanding geometry, we demonstrate 
the realization of the axial anomaly on the real-time lattice.
\end{abstract}

\maketitle

\section{Introduction} \label{sec:intro}
In relativistic heavy-ion collisions, CP-violating configurations of color gauge fields can be generated locally either by gauge field dynamics at the instant of a collision or sphaleron transitions at later times \cite{Kharzeev:2001ev,Lappi:2006fp,Lappi:2017skr,Moore:2010jd,Mace:2016svc}. 
Quarks interacting with such gauge fields induce the imbalance of axial charge due to the quantum phenomenon of axial anomaly.
In presence of a strong U(1) magnetic field, which may be generated in off-central collisions, the axial charge asymmetry can be converted to a flow of electric current along the magnetic field \cite{Vilenkin:1980fu}. 
This phenomenon
is called chiral magnetic effect (CME) \cite{Kharzeev:2007jp,Fukushima:2008xe,Kharzeev:2015znc}. 
Experimental searches for this novel phenomenon have been carried out at RHIC and the LHC \cite{Abelev:2009ac,Adamczyk:2014mzf,Abelev:2012pa}, where a charge dependence of azimuthal correlations was measured \cite{Voloshin:2004vk}. However, the observation of the CME in heavy-ion collisions still remains inconclusive due to large backgrounds \cite{Khachatryan:2016got}. 

On the theory side, there have been numerous developments in the description of the transport phenomena associated with the CME based on the chiral kinetic theories \cite{Stephanov:2012ki,Gao:2012ix,Chen:2012ca,Son:2012zy,Mueller:2017arw,Huang:2017tsq} and the anomalous hydrodynamics \cite{Newman:2005hd,Son:2009tf,Sadofyev:2010pr,Hirono:2014oda,Yin:2015fca,Shi:2017cpu}. 
To make predictions of observable consequences of the CME, some of these frameworks need the information of the axial charge distribution as an initial condition as well as the space-time distribution of the magnetic field.
Since the lifetime of the magnetic field is expected to be short $\ltsim$1 fm$/c$ \cite{Skokov:2009qp,Deng:2012pc,Tuchin:2013apa}, the understanding of the axial charge production at the early stage of heavy-ion collisions is indispensable in order to make reliable predictions about the CME. 

At high energies, colliding heavy-ions can be described in terms of the effective theory of color glass condensate (CGC) \cite{Iancu:2003xm,Weigert:2005us,Gelis:2010nm}. By a collision, strong color electromagnetic fields are generated, and the system expands to the longitudinal direction in a nearly boost-invariant way. 
Even though the QCD coupling is weak $g\ll 1$, this system, called glasma \cite{Lappi:2006fp}, is strongly correlated because the gauge fields are inversely proportional to the coupling constant $A\sim 1/g$ as a consequence of the gluon saturation. 
Nonperturbative dynamics of these gauge fields can be computed by classical(-statistical) gauge field simulations on the real-time lattice for the longitudinally expanding geometry \cite{Krasnitz:1998ns,Krasnitz:2001qu,Lappi:2003bi,Berges:2013eia,Berges:2013fga,Gelis:2013rba}. 
A key feature of the glasma is nonzero topological charge density $F\widetilde{F}$, which is comprised of longitudinal color electric and color magnetic fields having flux tube-like structures.\footnote{%
The physical picture of the glasma flux tube is similar to the flux tube model that is encoded in the Lund Monte Carlo model \cite{Andersson:1983ia}. 
One important difference of the glasma flux tube from the conventional color flux tube is the existence of color magnetic fields \cite{Lappi:2006fp}.
Besides it, a significant difference is the strength of color sources that generate the flux tubes. 
In the conventional flux tube picture, the color  source is a single pair of partons, which have an elementary charge of the order of $g$. In this case, once other single pair of partons is created in the flux tube via the Schwinger mechanism, the electric field is immediately shielded and string breaking happens. 
By contrast, the color source of the glasma flux tube is high-density gluons whose number density is $\sim 1/g$.
Since the charge density of this source is order one, a single pair of quark-antiquarks or gluons is not sufficient to shield the color field in the glasma. 
Consequently, the decay of the color field is not as sudden as the string-breaking picture. The field is gradually diluted as many pairs of particles are produced.
Meanwhile, the produced particles can coherently interact with the residual color field and the collective motion of the produced particles may appear \cite{Asakawa:1990se,Cooper:1992hw,Tanji:2010eu}.}
Because the typical field strength of the glasma is characterized by the saturation scale $Q_s$, which is much larger than the light-quark masses, quark production can happen intensely in the glasma.
Therefore, the glasma has the capability to generate abundant axial charges through the quark production.

Since the axial anomaly is a genuine quantum phenomenon, one needs to solve the dynamics of quantum quark fields for a proper description of the axial charge production. Once we approximate the time evolution of the strong gauge fields as that of classical fields, the dynamics of the quark fields under the strong gauge fields can be computed on the real-time lattice \cite{Gelis:2005pb,Gelis:2015eua,Tanji:2015ata,Gelfand:2016prm,Tanji:2016dka,%
Mueller:2016ven,Mace:2016shq,Tanji:2017xiw}. 
To the leading order in the coupling $g$, the quark dynamics under the strong gauge field $A\sim 1/g$ is governed by the Dirac equation that nonperturbatively couples to the gauge field via the covariant derivative \cite{Gelis:2015kya}. 
To this order of the approximation, the backreaction from quarks to the gauge field is negligible and the gauge field can be regarded as a background field. In the next-to-leading order of the weak-coupling and strong-field approximation, the Yang--Mills equations couple to current  induced by the quarks representing the effect of the backreaction \cite{Kasper:2014uaa}. 
In this study, we consider the weak-coupling and strong-field limit and thus neglect the backreaction.

The aim of this paper is to present formulation and numerical results for the axial charge production in the glasma gauge fields taking the expanding geometry specific to the early stage of heavy-ion collisions into account. 
To manifest the axial anomaly on a lattice, one has to take care of the fermion doubling problem \cite{Nielsen:1980rz}. 
In the context of the real-time lattice simulations, the Wilson fermion method has been successfully applied to the description of the axial anomaly in nonexpanding systems \cite{Aarts:1998td,Saffin:2011kn,Buividovich:2015jfa,Tanji:2016dka,Mueller:2016aao,Mueller:2016ven,Mace:2016shq}. 
We will employ the Wilson fermion method that is extended to the expanding geometry, which was first introduced in Ref.~\cite{Tanji:2017xiw}. 
Since the glasma gauge fields are produced as a consequence of the interactions between colliding two sheets of CGC, it is important for consistency to take the interactions of the quark fields with the CGC fields into account, \textit{i.e.} to solve the Dirac equation under the CGC gauge fields. 
The Dirac equation is analytically solvable until the time right after a collision \cite{Gelis:2005pb}, and the solution that explicitly manifests the boost invariance of the system has been derived in Ref.~\cite{Gelis:2015eua}. 
We will employ this solution as an initial condition for the time evolution after a collision.

In the framework of the CGC, classical gauge fields are emitted from color charges that represent hard degrees of freedom in a nucleus, and the distributions of the color charges are treated as random variables \cite{Iancu:2003xm,Weigert:2005us,Gelis:2010nm}. 
When two nuclei collide, the longitudinal color fields are generated depending on the color charges of each nucleus.
Since the two nuclei are causally separated before the collision and their color distributions are random, also the topological charge density $F\widetilde{F}$ has random nature: it fluctuates event by event, and in each event it has a random distribution in the transverse plane. 
As a first step to elucidate the axial charge production in the early stage of heavy-ion collisions, instead of the random color distributions, we consider fixed configurations of the color charges that realize simple flux tube-like  configurations of color fields that has nonzero $F\widetilde{F}$. 
By this setup, we aim at simulating the axial charge production in a domain where $F\widetilde{F}$ happens to be nonzero in a single collision event. 

The paper is organized as follows.
In Sec.~\ref{sec:Bjorken}, the formulation of the Dirac field and the axial anomaly in the boost-invariantly expanding system is explained.
In Sec.~\ref{sec:CGC}, we first review the CGC initial conditions for the gauge fields and the quarks fields, and then we explicitly construct the color charge distribution that realizes the flux-tube structure of the glasma color field. 
After we discuss the formulation of the problem on the real-time lattice in Sec.~\ref{sec:lattice}, we present our numerical results in Sec.~\ref{sec:result}.
First, we consider uniform glasma fields by taking the limit of large flux-tube width and verify that the axial anomaly is correctly realized on the real-time lattice in the expanding geometry. Then we show results for the glasma flux tube configuration. Section \ref{sec:conclusion} is devoted to concluding remarks.

In this paper, we use the metric $g^{\mu\nu} =\text{diag} (1,-1,-1,-1)$ in the original $(t,x,y,z)$ coordinates. 

\section{Axial anomaly in the Bjorken frame} \label{sec:Bjorken}
In the high energy limit of a heavy-ion collision, the system right after the collision shows boost-invariant expansion to the longitudinal direction, which can be conveniently described in terms of proper time $\tau$ and space-time rapidity $\eta$ defined by
\begin{equation}
\tau = \sqrt{t^2 -z^2} \, , \hspace{10pt}
\eta = \frac{1}{2} \ln \left( \frac{t+z}{t-z} \right) \, ,
\end{equation}
as well as transverse coordinates $\bxp =(x,y)$.

In the original rest frame, the vacuum expectation of the axial current density is given by
\begin{equation}
j_5^\mu (x) = \langle 0| \overline{\Psi} (x) \gamma^\mu \gamma_5 \Psi (x) |0\rangle \, ,
\end{equation}
where $\Psi (x)$ denotes the quark field operator.
The axial current obeys the Adler--Bell--Jackiw \cite{Adler:1969gk,Bell:1969ts} anomaly equation
\begin{equation}
\partial_\mu j_5^\mu =2m\langle 0|\overline{\Psi} i\gamma_5 \Psi |0\rangle 
+\frac{g^2}{4\pi^2} \bE^a \cd \bB^a \, ,
\label{ABJ0}
\end{equation}
where $m$ denotes quark mass and the summation over the color indices $a=1,\cdots ,N_c^2-1$ is implied.\footnote{%
In this paper, we consider only one quark flavor. Since we neglect the backreaction, different flavors contribute to the axial anomaly just additively.}

To respect the boost invariance, we compute all expectation values in the Bjorken frame that moves to the longitudinal direction with the local velocity of $v_z=z/t=\tanh \eta$. 
Moving to the Bjorken frame, the axial current is transformed as
\begin{equation}
\widehat{j}_5^\mu (x) = \Lambda^\mu_{\ \nu} j_5^\nu (x) \, ,
\label{j5_Bjo0}
\end{equation}
where 
\begin{equation}
\Lambda^\mu_{\ \nu}  = 
\begin{pmatrix}
\cosh \eta & 0 & 0 & -\sinh \eta \\ 0 & 1 & 0 & 0 \\ 0 & 0 & 1 & 0 \\ -\sinh \eta & 0 & 0 & \cosh \eta 
\end{pmatrix}
\end{equation}
is the boost operator to the Bjorken frame for four-vectors. Here and in the following, quantities in the Bjorken frame are denoted with a hat  $\ \widehat{}\ $. 
To solve the Dirac equation under boost-invariant background fields, it is convenient to treat the quark field operator  boosted to the Bjorken frame,
\begin{equation}
\widehat{\Psi} = \sqrt{\tau} e^{-\tfrac{\eta}{2} \gamma^0 \gamma^3} \Psi \, ,
\end{equation}
where $e^{-\tfrac{\eta}{2} \gamma^0 \gamma^3}$ is the boost operator to the Bjorken frame for spinors. 
The factor $\sqrt{\tau}$ is just a convention to make the following equation simpler. The Dirac equation for the boosted field is 
\begin{equation}
\left( i\gamma^0 D_\tau +\frac{i}{\tau} \gamma^3 D_\eta +i\gamma^i D_i -m \right) \widehat{\Psi} (x) = 0 \, ,
\label{Dirac0}
\end{equation}
with $D_\mu = \partial_\mu +igA_\mu $ being the covariant derivative \cite{Gelis:2015eua}.
Here and in the following, repeated indices $i$ imply the summation over $i=1,2$.
In terms of the boosted field operator, the axial current in the Bjorken frame \eqref{j5_Bjo0} is simply rewritten as
\begin{equation}
\widehat{j}_5^\mu (x) = \frac{1}{\tau} \langle 0| \overline{\widehat{\Psi}} (x) \gamma^\mu \gamma_5 \widehat{\Psi} (x) |0\rangle \, . 
\label{j5_Bjo}
\end{equation}
Such expectations of fermion operators can be expressed by fermion mode functions \cite{Aarts:1998td}. The mode functions are introduced by the mode expansion of the field operator
\begin{equation}
\widehat{\Psi} (\tau ,\eta ,\bxp ) 
= \sum_{s,c} \int \! \frac{d^2 p_{\!\perp} d\nu}{(2\pi)^3}
\left[ \widehat{\psi}_{\bpp ,\nu ,s,c}^+ (\tau ,\eta ,\bxp ) a_{\bpp ,\nu ,s,c}
+\widehat{\psi}_{\bpp ,\nu ,s,c}^- (\tau ,\eta ,\bxp ) b_{\bpp ,\nu ,s,c}^\dagger \right] \, ,
\label{mode}
\end{equation} 
where $a_{\bpp ,\nu ,s,c}$ and $b_{\bpp ,\nu ,s,c}$ are annihilation operators of a quark and an antiquark, respectively, having momentum $(\bpp ,\nu)$, which is conjugate to $(\bxp ,\eta)$, spin $s$ and color $c$. 
The superscripts $+$ and $-$ in the mode functions distinguish the positive and the negative energy solutions. 
By substituting Eq.~\eqref{mode} into \eqref{j5_Bjo}, we find the expression
\begin{equation}
\widehat{j}_5^\mu = \frac{1}{\tau} \sum_{s,c} \int \! \frac{d^2 p_{\!\perp} d\nu}{(2\pi)^3} \,
\overline{\widehat{\psi}}{}_{\bpp ,\nu ,s,c}^- \gamma^\mu \gamma_5 \widehat{\psi}_{\bpp ,\nu ,s,c}^- \, .
\end{equation}

In terms of the quantities in the Bjorken frame, the anomaly equation \eqref{ABJ0} is rewritten as
\begin{equation}
\frac{1}{\tau} \partial_\tau \left( \tau \widehat{j}_5^0 \right) +\partial_i \widehat{j}_5^i +\frac{1}{\tau} \partial_\eta \widehat{j}_5^3 = \frac{2m}{\tau} \langle 0| \overline{\widehat{\Psi}} i\gamma_5 \widehat{\Psi} |0\rangle +\frac{g^2}{4\pi^2} \bE^a \cd \bB^a \, .
\label{ABJ}
\end{equation}
We note that the two terms in the right hand side are Lorentz scalars. 
In boost-invariant background fields we consider in this study, the $\eta$-derivative term drops.
The axial charge density per unit transverse area and unit space-time rapidity is related with $\widehat{j}_5^0$ as
\begin{equation}
\frac{dN_5}{d^2 x_{\!\perp} d\eta} = \tau \widehat{j}_5^0 \, .
\end{equation}
By integrating Eq.~\eqref{ABJ} over the proper time, we find the relation
\begin{equation}
\frac{dN_5}{d^2 x_{\!\perp} d\eta} +\int_0^\tau \! \tau^\pr \partial_i \widehat{j}_5^i (\tau^\pr ,\bxp ) d\tau^\pr
= 2m \int_0^\tau \! \bar{\eta} (\tau^\pr ,\bxp ) d\tau^\pr
+\frac{g^2}{4\pi^2} \int_0^\tau \! \tau^\pr \bE^a \cd \bB^a d\tau^\pr \, ,
\label{budget}
\end{equation}
where we have introduced a shorthand notation for the pseudoscalar condensate,
\begin{equation}
\bar{\eta} = \langle 0| \overline{\widehat{\Psi}} i\gamma_5 \widehat{\Psi} |0\rangle \, .
\end{equation}
In deriving Eq.~\eqref{budget}, we have assumed that the axial charge density is vanishing at $\tau=0$, which is the case for the CGC initial condition discussed in the next section.

\section{CGC initial conditions} \label{sec:CGC}
In the CGC effective theory, hard degrees of freedom in a high energy nucleus are treated as classical sources of radiation, while soft degrees of freedom are described as classical gauge fields that couple to the hard sources via the Yang--Mills equations
\begin{equation}
\left[ D_\mu , F^{\mu \nu} \right] = J^\nu \, .
\label{YM} 
\end{equation}
The classical sources of two colliding nuclei running with the speed of light are represented by a current
\begin{equation}
J^\mu = \delta^{\mu +} \delta (x^- ) \rho_{(1)} (\bxp ) +\delta^{\mu -} \delta (x^+ ) \rho_{(2)} (\bxp ) \, ,
\end{equation}
where $\rho_{(n)} (\bxp )$ $(n=1,2)$ denote the color charge densities of the two nuclei in the transverse plane, 
and $x^\pm$ are light-cone coordinates defined by $x^\pm =(t\pm z)/\sqrt{2}$. 
With the initial condition $A^\mu =0$ at $t\to -\infty$, the Yang--Mills equations can be solved analytically up to the $\tau=0^+$ surface\footnote{By $\tau=0^+$, we denote an infinitesimal positive $\tau$.} \cite{Kovner:1995ja}. 
The solution at $\tau=0^+$ in the Fock--Schwinger gauge $A_\tau = 0$ is
\begin{align}
A^i (\tau=0, \bxp ) &= \alpha_{(1)}^i (\bxp ) +\alpha_{(2)}^i (\bxp ) \, , \label{Ai_t0} \\
A^\eta (\tau=0, \bxp ) &= \frac{ig}{2} \left[ \alpha_{(1)}^i ,\alpha_{(2)}^i \right] \, , \label{Aeta_t0}
\end{align}
where $\alpha_{(n)}^i$ are transverse pure gauges
\begin{align}
\alpha_{(n)}^i (\bxp ) &= -\frac{i}{g} V_{(n)}^\dagger (\bxp ) \partial^i V_{(n)} (\bxp ) \label{alpha}
\end{align}
associated with gauge factors
\begin{equation}
V_{(n)} (\bxp ) = \exp \left[ ig \nabla_{\!\perp}^{-2} \rho_{(n)} (\bxp ) \right]  \, .
\label{V_n}
\end{equation}
Nonzero components of the color electromagnetic fields given by these gauge fields are only longitudinal ones,
\begin{align}
E_z (\tau =0,  \bxp) &= -ig \left[ \alpha_{(1)}^i ,\alpha_{(2)}^i \right] \, , \label{Ez} \\
B_z (\tau=0, \bxp ) &= -ig \epsilon_{ij} \left[ \alpha_{(1)}^i ,\alpha_{(2)}^j \right] \, . \label{Bz}
\end{align}
This color field configuration in general carries nonzero topological charge density $\bE^a \cd \bB^a$, and hence can generate axial charge through quark production \cite{Lappi:2006fp}. We emphasize that nonzero $\bE^a \cd \bB^a$ exists only after a collision. The CGC color fields localized on the light cones, $x^\pm =0$, are only transverse ones and electric and magnetic fields are orthogonal to each other, $\bE \perp \bB$ \cite{Lappi:2006fp}. Therefore, the axial charge density is vanishing at the instant of a collision, $\tau=0$. 

In the McLerran--Venugopalan (MV) model \cite{McLerran:1993ni}, the color charges $\rho_{(n)} (\bxp )$ are assumed to be distributed randomly in the transverse plane according to a Gaussian probability distribution. 
In the present study, we consider a fixed configuration of $\rho_{(n)} (\bxp )$ that corresponds to a flux tube-like configuration of the color electromagnetic fields in order to elucidate the nonequilibrium axial charge production in a simpler situation.

The leading order dynamics of fermions under strong gauge fields can be described by the Dirac equation for the fermion mode functions \cite{Aarts:1998td,Gelis:2015kya}. 
Under the CGC gauge fields, the Dirac equation can be solved analytically up to the $\tau =0^+$ surface \cite{Gelis:2005pb,Gelis:2015eua}. 
The mode solution at $\tau=0^+$ with the initial condition of the negative-energy free spinor at $t\to -\infty$ is 
\begin{align}
\widehat\psi_{\bpp ,\nu ,s,a}^-(x)
&\underset{\tau=0^+}{=}
-\frac{e^{\frac{\pi}{4}i}}{\sqrt{4\pi M_\bp}}
e^{i\bpp \cdot \bxp +i\nu\eta} \int\frac{d^2 q_{\!\perp}}{(2\pi)^2} \frac{e^{i\bqp\cdot \bxp}}{M_{\bp +\bq}} \notag \\
&\hspace{10pt}
\times
\Bigg\{
e^{\frac{\pi\nu}{2}}\Big(\tfrac{M_{\bp +\bq}^2\tau}{2M_\bp}\Big)^{i\nu}
\Gamma(-i\nu +\tfrac{1}{2}) V_2^\dagger(\bxp) \widetilde{V}_2(\bqp ) \gamma^+
\notag \\
&\hspace{10pt}
+
e^{-\frac{\pi\nu}{2}}\Big(\tfrac{M_{\bp +\bq}^2\tau}{2M_\bp}\Big)^{-i\nu}
\Gamma(i\nu +\tfrac{1}{2})
V_1^\dagger(\bxp) \widetilde{V}_1(\bqp ) \gamma^- \Bigg\}
(q^i\gamma^i -M_\bp \gamma^0 )v_s(-\bpp ) \chi_a \, ,
\label{qini0}
\end{align}
where $M_\bp=\sqrt{m^2 +\bp_\perp^2}$ is the transverse mass, $v_s (\bp )$ is the negative-energy free spinor, and $\chi_a$ $(a=1,\cdots ,N_c)$ are unit vectors in the color space \cite{Gelis:2015eua}.\footnote{%
This expression is slightly changed from that given in Ref.~\cite{Gelis:2015eua}; the sign of the transverse momentum index is flipped, $\bpp \leftrightarrow -\bpp$, the integration variables are shifted as $\bqp \to \bqp +\bpp$, and the overall normalizations differ by the factor $\sqrt{4\pi}$.}
The Fourier transform of the gauge factors $\widetilde{V}_n (\bpp)$ are defined as
\begin{equation}
\widetilde{V}_n (\bpp) 
= \int \! d^2 x_{\!\perp} \, V_n (\bxp ) e^{-i\bpp \cdot \bxp} \, .
\end{equation}
For the spinor satisfying $v_s^\dagger  (\bp ) v_{s^\prime} (\bp^\prime ) =2\sqrt{\bp^2 +m^2} \, \delta_{s s^\prime}$, the mode functions are normalized as
\begin{align}
\int \! d^2 x_{\!\perp} d\eta \, \widehat\psi_{\bpp ,\nu ,s,a}^{-\, \dagger} (\tau ,\bxp ,\eta) 
\widehat\psi_{\bpp^\prime ,\nu^\prime ,s^\prime ,a^\prime}^- (\tau ,\bxp ,\eta) 
&= (2\pi)^3 \delta^2 (\bpp -\bpp^\prime ) \delta (\nu -\nu^\prime ) \delta_{ss^\prime} \delta_{aa^\prime} \, .
\end{align}

\subsection{Glasma flux tube} \label{subsec:IC_flux}
We will construct the gauge factors $V_{(n)}$ such that the initial longitudinal electric and magnetic fields have localized $\bxp$ dependences. In the following, we consider the color SU(2) theory for simplicity.

We suppose that each of the color sources $\rho_{(n)} (\bxp )$ has only one color component, and write
\begin{align}
V_{(1)} (\bxp ) = \exp \left[ i\Theta_1 (\bxp ) \frac{\sigma^1}{2} \right] \, , \hspace{10pt}
V_{(2)} (\bxp ) = \exp \left[ i\Theta_2 (\bxp ) \frac{\sigma^2}{2} \right] \label{Vn}
\end{align}
with real functions $\Theta_n (\bxp )$ and the Pauli matrices $\sigma^i$. 
For these $V_{(n)}$, the transverse gauge fields $\alpha_{(n)}^i$ are expressed as
\begin{equation}
\alpha_{(n)}^i  (\bxp ) = -\frac{1}{g} \partial_i \Theta_n (\bxp ) \frac{\sigma^n}{2} \, .
\end{equation}
Since the color orientations of $\alpha_{(1)}$ and $\alpha_{(2)}$ are different, the electric and magnetic fields right after the collision given by \eqref{Ez} and \eqref{Bz} can be nonzero for these configurations. 
In a realistic situation of a heavy-ion collision, the color charges $\rho_{(n)} (\bxp )$ are distributed randomly in the transverse plane. In that case, the color orientations of the pure gauges $\alpha_{(n)}$ at a certain $\bxp$ should be given randomly, and the value of $\bE^a \cd \bB^a$ right after the collision fluctuates in the transverse plane.
In Eqs.~\eqref{Vn}, we have chosen one of specific color configurations that realize nonzero $\bE^a \cd \bB^a$ right after the collision. 
We note that the color configurations of each nucleus, Eqs.~\eqref{Vn}, should be understood as ones given in a common gauge that is globally fixed. Until the instant of collision, the color orientations of each nucleus do not have any physical meaning because the two nuclei are causally separated and one can apply independent gauge transformations to them. However, it is not the case anymore after the collision. Since the color fields after the collision are generated by the interaction between the two nuclei, they should be computed in a common gauge. 

We further assume that the functions $\Theta_n (\bxp )$ depend on $x$ and $y$ only through the combination of $\xi_n = x\cos \theta_n +y\sin \theta_n$ with real parameters $\theta_n$.
Then, $\alpha_{(n)}^i$ can be written as
\begin{equation} 
\alpha_{(n)}^1  (\bxp ) = -\frac{1}{g} \mathcal{Q}_n (\xi_n ) \cos \theta_n \frac{\sigma^n}{2} \,  , \hspace{10pt}
\alpha_{(n)}^2  (\bxp ) = -\frac{1}{g} \mathcal{Q}_n (\xi_n ) \sin \theta_n \frac{\sigma^n}{2} \, 
\label{alpha12}
\end{equation}
where we have introduced
\begin{equation}
\mathcal{Q}_n (\xi_n ) = \frac{\partial}{\partial \xi_n} \Theta_n (\xi_n ) \, .
\end{equation}
For these $\alpha_{(n)}^i $, the initial electric and magnetic fields read
\begin{align}
E_z (\tau =0,  \bxp) 
&= \frac{1}{g} \mathcal{Q}_1 \mathcal{Q}_2 \cos (\theta_1 -\theta_2 ) \frac{\sigma^3}{2} \, , \\
B_z (\tau=0, \bxp ) 
&= -\frac{1}{g} \mathcal{Q}_1 \mathcal{Q}_2 \sin (\theta_1 -\theta_2 ) \frac{\sigma^3}{2} \, .
\end{align}
The topological charge density $\bE^a \cd \bB^a$ is nonzero when $\theta_1 -\theta_2 \neq \pi n/2$ ($n$: integers).

To gain a flux tube-like structure with a Gaussian profile for the electric and magnetic fields,  we assume
\begin{equation}
\Theta_n (\xi_n ) 
= \frac{\sqrt{\pi}}{2} Q_n \Delta \, \text{Erf} \left( \frac{\xi_n}{\Delta} \right) \, , 
\label{Theta0}
\end{equation}
where $Q_n$ are parameters that have the mass dimension one, $\Delta$ characterizes the width of a flux tube, and
$\text{Erf} (x)$ is the error function.
This leads
\begin{equation}
\mathcal{Q}_n (\xi_n )
= Q_n \exp \left( -\frac{\xi_n^2}{\Delta^2} \right) \, .
\end{equation}
Then, the $\bxp$ dependence of the electric and magnetic fields turns out to be a distorted Gaussian\footnote{%
Although the electric and magnetic fields are localized in the transverse plane (unless $\cos (\theta_1 -\theta_2 )=\pm 1$), the color charges $\rho_{(n)}$ that generate these fields have infinitely elongated structures to the directions along $(x,y)=(\sin \theta_n ,-\cos \theta_n )$:
\begin{equation}
\rho_{(n)} (\bxp ) = -\frac{2Q_n}{g} \frac{\xi_n}{\Delta^2} \exp \left( -\frac{\xi_n^2}{\Delta^2} \right) \frac{\sigma^n}{2} \, .
\end{equation}
The electric and magnetic fields are induced in the overlapped region of the color charges $\rho_{(1)}$ and $\rho_{(2)}$. 
Note that these color charge distributions are globally color neutral, $\int \!d^2 x_{\!\perp} \, \rho_{(n)} (\bxp) =0$. 
This condition is necessary for the inverse Laplacian in Eq.~\eqref{V_n} being well-defined. 
},
\begin{align}
&\mathcal{Q}_1 (\xi_1 ) \mathcal{Q}_2 (\xi_2 ) \notag \\
&= Q_1 Q_2 \exp \left[ -\frac{x^2+y^2+(x^2-y^2)\cos (\theta_1+\theta_2) \cos (\theta_1-\theta_2) +2xy\sin (\theta_1+\theta_2) \cos (\theta_1-\theta_2)}{\Delta^2} \right] \, .
\end{align}

\subsection{Uniform glasma} \label{subsec:IC_uniform}
By taking the limit of an infinitely wide flux tube, $\Delta \to \infty$, we obtain a uniform color field configuration that carry nonzero topological charge. 
In this limit, the functions $\Theta_n (\xi_n)$ become linear in $\xi_n$,
\begin{equation}
\Theta_n (\xi_n ) 
= Q_n \xi_n \, , 
\end{equation}
and the electric and magnetic fields become uniform,
\begin{align}
E_z (\tau =0,  \bxp) 
&= \frac{1}{g} Q_1 Q_2 \cos (\theta_1 -\theta_2 ) \frac{\sigma^3}{2} \, , \\
B_z (\tau=0, \bxp ) 
&= -\frac{1}{g} Q_1 Q_2 \sin (\theta_1 -\theta_2 ) \frac{\sigma^3}{2} \, .
\end{align}

By substituting $V_{(n)} = \exp \left[ iQ_n \xi_n \tfrac{\sigma_n}{2} \right]$ into Eq.~\eqref{qini0}, we find the quark mode function at $\tau=0^+$ for this background,
\begin{align}
\widehat\psi_{\bpp ,\nu ,s,a}^-(x)
&\underset{\tau=0^+}{=}
\frac{e^{\frac{\pi}{4}i}}{\sqrt{4\pi M_\bp}}
e^{i\bpp \cdot \bxp +i\nu\eta} \notag \\
& \times \Bigg\{
\frac{1}{M_{\bp +\bq_2}} \left( \frac{M_{\bp +\bq_2}^2 \tau}{2M_\bp} \right)^{i\nu} e^{\frac{\pi\nu}{2}} 
\Gamma (-i\nu +\tfrac{1}{2} ) \frac{1+\sigma^2}{2} \gamma^+ \left( \gamma^0 M_\bp -\bgam \cdot \bq_2 \right) 
\notag \\
&\hspace{10pt} +
\frac{1}{M_{\bp -\bq_2}} \left( \frac{M_{\bp -\bq_2}^2 \tau}{2M_\bp} \right)^{i\nu} e^{\frac{\pi\nu}{2}} 
\Gamma (-i\nu +\tfrac{1}{2} ) \frac{1-\sigma^2}{2} \gamma^+ \left( \gamma^0 M_\bp +\bgam \cdot \bq_2 \right) 
\notag \\
&\hspace{10pt} +
\frac{1}{M_{\bp +\bq_1}} \left( \frac{M_{\bp +\bq_1}^2 \tau}{2M_\bp} \right)^{-i\nu} e^{-\frac{\pi\nu}{2}} 
\Gamma (i\nu +\tfrac{1}{2} ) \frac{1+\sigma^1}{2} \gamma^- \left( \gamma^0 M_\bp -\bgam \cdot \bq_1 \right) 
\notag \\
&\hspace{10pt} +
\frac{1}{M_{\bp -\bq_1}} \left( \frac{M_{\bp -\bq_1}^2 \tau}{2M_\bp} \right)^{-i\nu} e^{-\frac{\pi\nu}{2}} 
\Gamma (i\nu +\tfrac{1}{2} ) \frac{1-\sigma^1}{2} \gamma^- \left( \gamma^0 M_\bp +\bgam \cdot \bq_1 \right) 
\Bigg\} v_s (-\bpp ) \chi_a \, ,
\label{qini_uni}
\end{align}
where we have introduced two-dimensional vectors $\bq_n = (\tfrac{1}{2}Q_n \cos \theta_n ,\tfrac{1}{2}Q_n \sin \theta_n )$. 
By using this expression, one can directly confirm that the axial charge density $\widehat{j}_5^0$ is vanishing at $\tau=0^+$. 
Nevertheless this expression is indicative of axial charge imbalance that is induced right after $\tau=0^+$, as the $\gamma^+$ and $\gamma^-$ projections are asymmetric when $\bq_1 \neq \bq_2$. 

\section{Lattice formulation} \label{sec:lattice}
Since it is difficult to analytically solve the Dirac equation and the classical Yang--Mills equation for $\tau >0$ with the CGC initial conditions, we resort to numerical computations on the real-time lattice. 
The lattice discretization method we employ is the same as that used in Ref.~\cite{Tanji:2017xiw}. We will review it in this section to make the paper self-contained, and also explain issues specific to the present study. 
The space coordinates $(\bxp ,\eta )$ are discretized into $N_\perp \times N_\perp \times N_\eta$ grids with spacings $(\ap ,\ap ,a_\eta )$. The transverse and the longitudinal system size are $L_\perp =N_\perp \ap$ and $L_\eta =N_\eta a_\eta$, respectively. 
The periodic boundary condition is imposed on all the fields. 

\subsection{Gauge sector} \label{subsec:latt_gauge}
On the spatial lattice, the gauge fields are represented by link variables $U_i$, $U_\eta $ and electric fields $E^i$, $E^\eta$, where $i=1,2$ denotes the transverse directions. The link variables are related with the original gauge fields as
\begin{equation}
U_i (x) = \exp \left[ ig\ap A_i (x)\right] \, , \hspace{10pt}
U_\eta (x) = \exp \left[ iga_\eta A_\eta (x)\right] \, .
\end{equation}
The physical electric fields in the Bjorken frame are related with the lattice electric fields as $(\widehat{E}_x ,\widehat{E}_y ,\widehat{E}_z )=(E^1/\tau ,E^2/\tau ,E^\eta )$. 
The lattice version of the classical Yang--Mills equations in the expanding geometry are
\begin{align}
\partial_\tau U_i (x) = ig \frac{\ap}{\tau} E^i (x) U_i (x) 
\hspace{20pt} (\text{no summation over } i),
\label{lattYM1}
\end{align} 
\begin{align}
\partial_\tau U_\eta (x) = ig a_\eta \tau E^\eta (x) U_\eta (x) \, , \label{lattYM2}
\end{align}
\begin{align}
\partial_\tau E^i (x) 
&= -\frac{\tau}{g\ap^3} \sum_{j\neq i} \text{Im} \left[ U_{i,j} (x)+U_{i,-j} (x)\right]_\text{traceless} \notag \\
&\hspace{10pt} 
-\frac{1}{g\tau \ap a_\eta^2} \text{Im} \left[ U_{i,\eta} (x)+U_{i,-\eta} (x)\right]_\text{traceless} \, , 
\label{lattYM3}
\end{align}
and
\begin{align}
\partial_\tau E^\eta (x) 
&= -\frac{1}{g\tau a_\eta \ap^2} \sum_{i=1,2} \text{Im} \left[ U_{\eta ,i} (x)+U_{\eta ,-i} (x)\right]_\text{traceless} \, ,
\label{lattYM4}
\end{align}
where the subscript \lq{}traceless\rq{} means
\begin{equation}
\left[ X\right]_\text{traceless} = X -\frac{1}{N_c} \text{tr} \left( X\right) \, .
\end{equation}
The plaquettes variables $U_{\mu ,\nu} (x)$ and $U_{\mu ,-\nu} (x)$ are defined by
\begin{equation}
U_{\mu ,\nu} (x) = U_\mu (x) U_\nu (x+\hat{\mu}) U_\mu^\dagger (x+\hat{\nu}) U_\nu^\dagger (x) \, ,
\end{equation}
and
\begin{equation}
U_{\mu ,-\nu} (x) = U_\mu (x) U_\nu^\dagger (x+\hat{\mu} -\hat{\nu} ) U_\mu^\dagger (x-\hat{\nu} ) U_\nu (x-\hat{\nu}) \, ,
\end{equation}
with $\hat{\mu}$ representing the unit displacement in the $\mu$ direction on the lattice. 
To give a definition of magnetic fields, we further introduce other kinds of plaquettes,
\begin{align}
U_{-\mu ,\nu} (x) &= U_\mu^\dagger (x-\hat{\mu}) U_\nu (x-\hat{\mu}) U_\mu^\dagger (x-\hat{\mu}+\hat{\nu}) U_\nu^\dagger (x) \, , \\
U_{-\mu ,-\nu} (x) &= U_\nu^\dagger (x-\hat{\nu}) U_\mu^\dagger (x-\hat{\mu} -\hat{\nu} ) U_\nu (x-\hat{\mu}-\hat{\nu} ) U_\mu (x-\hat{\mu}) \, .
\end{align}
Then we adopt the four-plaquettes definition of the magnetic fields,
\begin{align}
B_i^a (x) &= \frac{-1}{2ga_\perp \tau a_\eta} \sum_{j\neq i} \text{Im} \, \text{tr}
\left[ T^a \left( U_{j,\eta} (x) +U_{\eta ,-j} (x) +U_{-\eta ,j} (x) +U_{-j,-\eta} (x) \right) \right] \, , \\
B_z^a (x) &= \frac{-1}{2ga_\perp^2} \text{Im} \, \text{tr}
\left[ T^a \left( U_{1,2} (x) +U_{2 ,-1} (x) +U_{-2 ,1} (x) +U_{-1,-2} (x) \right) \right] \, ,
\end{align}
where $T^a$ is the generators of SU($N_c$). 

\subsection{Quark sector} \label{subsec:latt_quark}
The Dirac equation for the mode functions on the lattice has almost the same form as that for the field operator in the continuum \eqref{Dirac0},
\begin{equation}
\left( i\gamma^0 \partial_\tau +\frac{i}{\tau} \gamma^3 D_\eta +i\gamma^i D_i -m +W \right) \widehat{\psi}_{\bpp ,\nu ,s,c}^- = 0 \, .
\label{Dirac}
\end{equation}
The differences are the form of the covariant derivative and the addition of the Wilson term. 
To improve the convergence to the continuum limit, we employ the $\mathcal{O} (a^3)$-improved lattice derivatives \cite{Mueller:2016ven,Mace:2016shq}. 
The covariant derivative is given by
\begin{align}
D_\mu \psi (x) 
&= \frac{c_1}{a_\mu} \left[ U_\mu (x) \psi (x+\hat{\mu} ) -U_\mu^\dagger (x-\hat{\mu} ) \psi (x-\hat{\mu} ) \right] \notag \\
&\hspace{10pt}
+\frac{c_2}{a_\mu} \left[ U_\mu (x) U_\mu (x+\hat{\mu} ) \psi (x+2\hat{\mu} ) 
-U_\mu^\dagger (x-\hat{\mu} ) U_\mu^\dagger (x-2\hat{\mu} ) \psi (x-2\hat{\mu} ) \right] \, ,
\end{align}
with coefficients $c_1=4/3$ and $c_2=-1/6$. 
As the spatial Wilson term extended to the expanding geometry, we employ
\begin{align}
W\psi (x) &= \frac{r_\perp}{2a_\perp} \sum_{i=1,2} \left\{ 
c_1 \left[ U_i(x) \psi (x+\hat{i}) -2\psi(x) +U_i^\dagger (x-\hat{i}) \psi (x-\hat{i} )\right] \right. \notag \\
&\hspace{10pt} \left. 
+2c_2\left[ U_i(x) U_i(x+\hat{i}) \psi (x+2\hat{i}) -2\psi(x) +U_i^\dagger (x-\hat{i}) U_i^\dagger (x-2\hat{i}) \psi (x-2\hat{i} )\right] \right\} \notag \\
&\hspace{10pt}
+\frac{r_\eta}{2T a_\eta} \left\{
c_1 \left[ U_\eta(x) \psi (x+\hat{\eta}) -2\psi(x) +U_\eta^\dagger (x-\hat{\eta}) \psi (x-\hat{\eta} )\right] \right. \notag \\
&\hspace{10pt} \left.
+2c_2 \left[ U_\eta(x) U_\eta(x+\hat{\eta}) \psi (x+2\hat{\eta}) -2\psi(x) +U_\eta^\dagger (x-\hat{\eta}) U_\eta^\dagger (x-2\hat{\eta}) \psi (x-2\hat{\eta} )\right] \right\} \, ,
\label{Wilson}
\end{align}
where $r_\perp$ and $r_\eta$ are real parameters, and $T$ is a quantity that has the dimension of time. 
In Ref.~\cite{Tanji:2017xiw}, $T=\tau_0$ (initial time) was employed. In the present study, we use $T=\tau$, which is essential to compute the early-time behavior of the axial charge production.

\begin{figure}[tb]
 \begin{center}
  \includegraphics[clip,width=8cm]{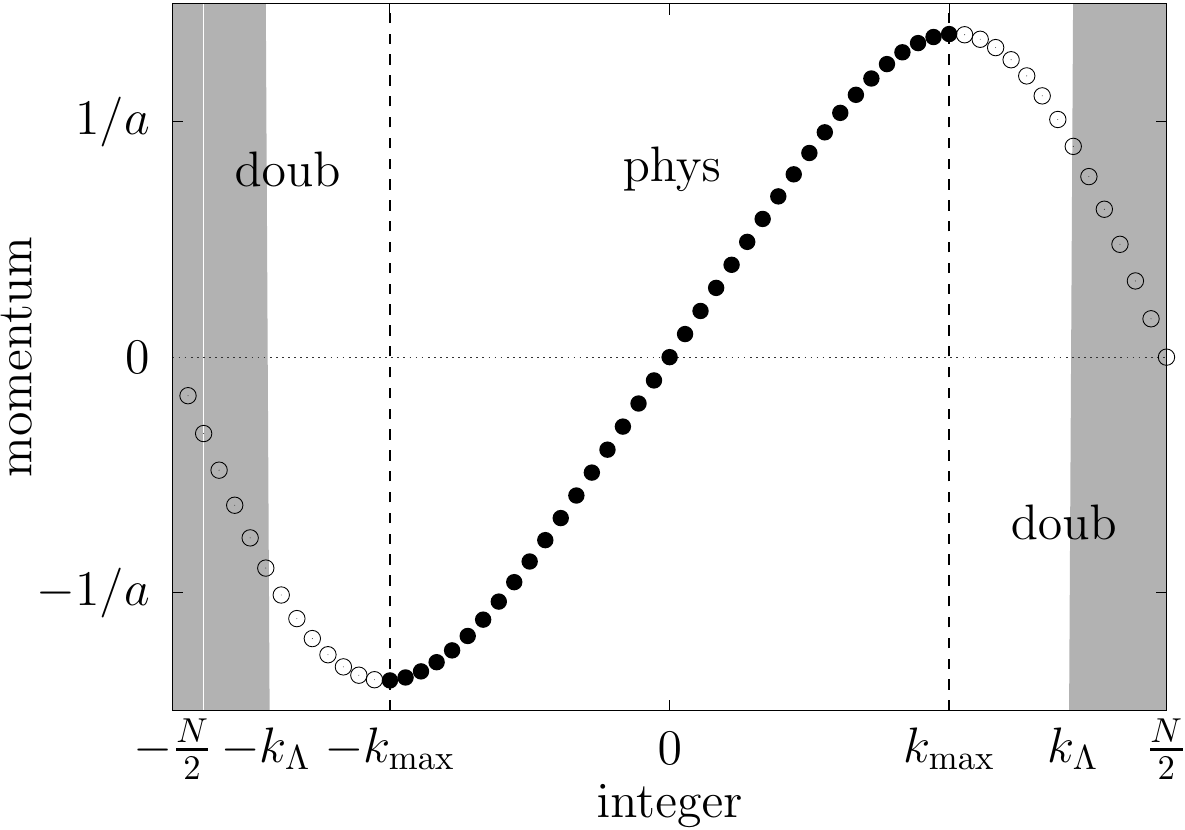} 
  \vspace{-8pt}
  \caption{A schematic plot of the fermion lattice momenta \eqref{lattmom_xy} and \eqref{lattmom_eta}. The physical modes and the doubler modes are separated at the integers $\pm k_\text{max}$, where $k_\text{max}\approx 0.286 N$. 
Modes corresponding to integers $|k|>k_\Lambda$ (gray shaded area) are excluded from computations to reduce the numerical cost.}
  \label{fig:lattmom}
 \end{center}
\end{figure}

On the lattice with the periodic boundary condition, the plane wave factor is replaced as
\begin{equation}
e^{i\bpp \cdot \bxp +i\nu \eta} \longrightarrow 
\exp \left[ 2\pi i \left( \frac{k_x n_x}{N_\perp} +\frac{k_y n_y}{N_\perp} +\frac{k_\nu n_\eta}{N_\eta} \right) \right] \, ,
\label{latt_pw}
\end{equation}
where $n_{x,y,\eta}$ are integers for the space coordinates; $(x,y,\eta ) = (\ap n_x ,\ap n_y ,a_\eta n_\eta )$, while integers $k_{x,y,\nu}$ specify the momentum modes. By the latter integers, fermion momenta are discretized as
\begin{equation}
p_{x,y} = \frac{c_1}{\ap} \sin \left( 2\pi \frac{k_{x,y}}{N_\perp} \right) 
+\frac{c_2}{\ap} \sin \left( 4\pi \frac{k_{x,y}}{N_\perp} \right) \hspace{10pt}
(k_{x,y} =-N_\perp/2 +1 ,\cdots, 0, \cdots, N_\perp /2) \, ,
\label{lattmom_xy}
\end{equation}
and
\begin{equation}
\nu = \frac{c_1}{a_\eta} \sin \left( 2\pi \frac{k_\nu}{N_\eta} \right) 
+\frac{c_2}{a_\eta} \sin \left( 4\pi \frac{k_\nu}{N_\eta} \right) \hspace{10pt}
(k_\nu =-N_\eta/2 +1 ,\cdots, 0, \cdots, N_\eta /2) \, .
\label{lattmom_eta}
\end{equation}
The dispersion of these lattice momenta is illustrated in Fig.~\ref{fig:lattmom}. 
The regions for integers $k$ satisfying $|k| >k_\text{max}\approx 0.286N$ correspond to fermion doubler modes. 
In the Wilson fermion method, the doublers are decoupled from the physical modes being made heavy by the Wilson term.

For the initial conditions of the quark mode functions, we replace \eqref{qini0} by
\begin{align}
\widehat\psi_{\bpp ,\nu ,s,a}^-(x)
&=
\frac{1}{\sqrt{4\pi M_\bp}}
e^{2\pi i \left( k_x n_x /N_\perp +k_y n_y /N_\perp +k_\nu n_\eta /N_\eta \right)}
\frac{1}{L_\perp^2} \sum_{j_x ,j_y} \frac{1}{M_{\bp +\bq}} 
e^{2\pi i \left( j_x n_x /N_\perp +j_y n_y /N_\perp \right)}\notag \\
&\hspace{10pt}
\times
\Bigg\{
e^{\frac{\pi\nu}{2}}\left( \frac{M_{\bp +\bq}^2\tau}{2M_\bp}\right)^{i\nu}
\Gamma(-i\nu +\tfrac{1}{2}) V_2^\dagger(\bxp) \widetilde{V}_2(\bqp ) \gamma^+
\notag \\
&\hspace{10pt}
+
e^{-\frac{\pi\nu}{2}}\left( \frac{M_{\bp +\bq}^2\tau}{2M_\bp}\right)^{-i\nu}
\Gamma(i\nu\!+\!{\tfrac{1}{2}})
V_1^\dagger(\bxp) \widetilde{V}_1(\bqp ) \gamma^- \Bigg\}
(\bqp \cd \bgam -M_\bp \gamma^0 )v_s(-\bpp ) \chi_a \, ,
\label{qini_latt}
\end{align}
where momenta $\bqp =(q_x ,q_y)$ are associated with integers $(j_x ,j_y)$. The transverse masses $M_\bp$ and $M_{\bp +\bq}$ in the above express contain the contribution from the Wilson term,
\begin{equation}
M_\bp = \sqrt{(m +m_W)^2 +\bpp^2} \, , \hspace{10pt}
M_{\bp +\bq} = \sqrt{(m +m_W)^2 +(\bpp +\bqp)^2} \, ,
\end{equation}
where the Wilson mass $m_W$ depends on $k_{x,y,\nu}$ and $\tau$ as
\begin{equation}
m_W = \sum_{l=1,2} c_l 
\left[ \frac{2r_\perp}{\ap} \sin^2 \left( l\pi \frac{k_x}{N_\perp} \right) +\frac{2r_\perp}{\ap}  \sin^2 \left( l\pi \frac{k_y}{N_\perp} \right) +\frac{2r_\eta}{\tau a_\eta} \sin^2 \left( l\pi \frac{k_\nu}{N_\eta} \right) \right] \, .
\end{equation}
The Fourier transform of the gauge factors is discretized as
\begin{equation}
\widetilde{V}_n (\bqp) 
= \ap^2 \sum_{n_x ,n_y} \, V_n (\bxp ) e^{-2\pi i \left( j_x n_x +j_y n_y \right) /N_\perp} \, .
\end{equation}

The most costly part in our numerical computation is solving the Dirac equation for the fermion mode functions. 
For general background gauge fields, the numerical cost to solve Eq.~\eqref{Dirac} is proportional to the square of the lattice size, $(N_\perp^2 N_\eta)^2$, since we have to solve the equations for all the modes and each mode function has dependence on the space coordinates.
In the longitudinally expanding system, the longitudinal lattice size $N_\eta$ especially needs to be large to resolve the longitudinal momentum scales that rapidly vary as $1/\tau$, and the numerical cost becomes unacceptably expensive.
One possible way to reduce the numerical cost is the use of the stochastic method for fermions \cite{Borsanyi:2008eu,Gelis:2015kya}. However, this method is not suitable to the computations of local quantities which are not averaged over space, especially quantities related with axial anomaly, because of large statistical errors. In this study, therefore, we stick to the direct method of solving the Dirac equation for the mode functions, which does not involve statistical errors.
Fortunately, the cost of the mode function method can be reduced by the factor of $N_\eta$ in boost-invariant backgrounds because the $\eta$-dependence of the mode functions is known to be $e^{i\nu \eta}$. 

To further reduce the numerical cost, we introduce cutoffs in the momentum space. As illustrated in Fig.~\ref{fig:lattmom}, the lattice fermion modes contain unphysical doubler modes. 
A naive way to regulate the doubler modes in the mode function method is just to cut off these modes from the computation. 
This approach has been successfully employed in Ref.~\cite{Tanji:2015ata} and also in Ref.~\cite{Tanji:2017xiw} being combined with the stochastic fermion method. 
However, this approach is not applicable to the computation of the axial anomaly because it amounts to introducing a cutoff for \textit{canonical} momentum and thus breaks the gauge invariance \cite{Tanji:2016dka}. 
Therefore, in the present study we employ the Wilson fermion, which amounts to introducing a cutoff for \textit{kinetic} momentum.
Once the doubler modes are suppressed by the Wilson term in a gauge-invariant way, we can introduce momentum cutoffs without affecting the axial anomaly. As depicted in Fig.~\ref{fig:lattmom}, we put a cutoff $k_\Lambda$ for the momentum integers between $k_\text{max}$ and $N/2$, and excluded the modes for $|k|>k_\Lambda$ from the computation. 
We have explicitly confirmed for the uniform glasma configuration that this cutoff does not alter the results for the axial charge production as long as $k_\Lambda$ is not too close to $k_\text{max}$. 
We typically choose a value of $k_\Lambda$ so that about 20\% of modes is excluded in each dimension. By this, the total numerical cost becomes about half ($0.8^3 \approx 0.5$).

\subsection{Axial anomaly on the lattice} \label{subsec:latt_anomaly}
On the lattice, the axial anomaly is a nontrivial issue. 
If one uses a naively discretized fermion action, degenerated doubler modes appear as shown in Fig.~\ref{fig:lattmom} and the axial anomaly is not realized due to the cancellation among the doublers \cite{Nielsen:1980rz}. For the axial anomaly, one needs to eliminate the doublers to spoil this cancellation.
As already discussed in the previous subsection, we employ the Wilson fermion method for this purpose.
The Wilson term \eqref{Wilson} introduced in the Dirac equation \eqref{Dirac} makes the doublers as heavy as the lattice ultraviolet (UV) cutoff scale and decouples them from the dynamics.
In the following, we explain how the axial anomaly is realized by the Wilson fermion.

Since we treat the time $\tau$ as a continuum variable, the definition of the time component of the axial current is the same as that in the continuum,
\begin{align}
\widehat{j}_5^0 (x) 
&= \frac{1}{\tau} \langle 0| \overline{\widehat{\Psi}} (x) \gamma^0 \gamma_5 \widehat{\Psi} (x) |0\rangle \notag \\
&= \frac{1}{\tau} \sum_{s,c} \frac{1}{L_\perp^2 L_\eta} \sum_{\bpp ,\nu} \,
\overline{\widehat{\psi}}{}_{\bpp ,\nu ,s,c}^- (x) \gamma^0 \gamma_5 \widehat{\psi}_{\bpp ,\nu ,s,c}^- (x) \, .
\end{align}
The spatial components must be modified on the lattice as \cite{Mace:2016shq}
\begin{align}
\widehat{j}_5^i (x) 
= \frac{1}{\tau} \sum_{s,c} \frac{1}{L_\perp^2 L_\eta} \sum_{\bpp ,\nu} \, &\Bigg\{
c_1 \text{Re} \left[ 
\overline{\widehat{\psi}}{}_{\bpp ,\nu ,s,c}^- (x) \gamma^i \gamma_5 U_i (x) \widehat{\psi}_{\bpp ,\nu ,s,c}^- (x+\hat{i}) \right] \notag \\
&
+c_2 \text{Re} \left[ 
\overline{\widehat{\psi}}{}_{\bpp ,\nu ,s,c}^- (x) \gamma^i \gamma_5 U_i (x) U_i (x+\hat{i}) \widehat{\psi}_{\bpp ,\nu ,s,c}^- (x+2\hat{i}) \right. \notag \\
&\hspace{30pt} \left. 
+\overline{\widehat{\psi}}{}_{\bpp ,\nu ,s,c}^- (x-\hat{i}) \gamma^i \gamma_5 U_i (x-\hat{i}) U_i (x) \widehat{\psi}_{\bpp ,\nu ,s,c}^- (x+\hat{i})  \right] \Big\} \, .
\end{align}
This expression is valid also for $i=3$ though the third component does not appear in the following equations due to the boost invariance. 
For this definition of the axial current, the anomaly equation \eqref{ABJ} is modified to
\begin{equation}
\frac{1}{\tau} \partial_\tau \left( \tau \widehat{j}_5^0 \right) +\nabla_{\!i}\, \widehat{j}_5^i 
= \frac{2m}{\tau} \bar{\eta} +\frac{g^2}{4\pi^2} \bE^a \cd \bB^a \, ,
\label{ABJ_latt}
\end{equation}
with $\nabla_{\!i}$ denoting the backward difference
\begin{equation}
\nabla_{\!i} \psi (x) = \frac{1}{a_i} \left[ \psi (x) -\psi (x-\hat{i})\right] \, .
\end{equation}
The pseudoscalar condensate $\bar{\eta} (x)$ is represented by the mode functions as
\begin{equation}
\bar{\eta} (x) = \sum_{s,c} \frac{1}{L_\perp^2 L_\eta} \sum_{\bpp ,\nu} \,
\overline{\widehat{\psi}}{}_{\bpp ,\nu ,s,c}^- (x) i\gamma_5 \widehat{\psi}_{\bpp ,\nu ,s,c}^- (x) \, .
\end{equation}

From the Dirac equation \eqref{Dirac} that includes the Wilson term, one can derive the relation
\begin{equation}
\frac{1}{\tau} \partial_\tau \left( \tau \widehat{j}_5^0 \right) +\nabla_{\!i}\, \widehat{j}_5^i 
= \frac{2m}{\tau} \bar{\eta} +w(x) \, ,
\end{equation}
where $w(x)$ stands for the expectation of a fermion operator involving $i\gamma_5$ and $W$,
\begin{align}
w(x) 
&= -\frac{2}{\tau} \text{Re} \sum_{s,c} \frac{1}{L_\perp^2 L_\eta} \sum_{\bpp ,\nu} \,
\overline{\widehat{\psi}}{}_{\bpp ,\nu ,s,c}^- (x) i\gamma_5 W \widehat{\psi}_{\bpp ,\nu ,s,c}^- (x) \, .
\label{Wilson_cont}
\end{align}
Comparing this equation with Eq.~\eqref{ABJ_latt}, we notice that the axial anomaly is realized by the Wilson fermion if
\begin{equation}
w (x) = \frac{g^2}{4\pi^2} \bE^a \cd \bB^a \, .
\end{equation}
This relation has been proven to hold in the continuum limit in the context of the Euclidean lattice gauge theory \cite{Karsten:1980wd,Rothe:1998ba}. In the context of the real-time lattice computations, it has been numerically confirmed for nonexpanding systems \cite{Tanji:2016dka,Mueller:2016aao,Mueller:2016ven,Mace:2016shq}. 

\subsection{Boundary condition} \label{subsec:bc}
As already noted, we impose the periodic boundary condition (b.c.) on the spatial lattice. 
For the flux tube configuration introduced in Sec.~\ref{subsec:IC_flux}, we need a special care for the periodicity in the transverse directions. 
In the gauge sector, the transverse link variables $U_i$ and the longitudinal electric field $E^\eta$ must satisfy the periodic b.c. at the initial time. Other components trivially satisfy the b.c. as they are vanishing then. 
Since the initial longitudinal electric field has a localized Gaussian profile, the field is vanishing at the boundaries and satisfies the periodic b.c. if the flux tube is located sufficiently away from the boundaries. In contrast, the initial transverse gauge fields
\begin{equation}
A^i (\tau =0, \bxp ) 
= -\frac{1}{g} \partial_i \Theta_1 (\bxp ) \frac{\sigma^1}{2} -\frac{1}{g} \partial_i \Theta_2 (\bxp ) \frac{\sigma^2}{2}
\end{equation}
have an elongated structure to the directions along $(x,y)=(\sin \theta_1 ,-\cos \theta_1)$ and $(\sin \theta_2 ,-\cos \theta_2)$ for the functions $\Theta_n (\bxp )$ given by \eqref{Theta0}. 
On a square transverse lattice and for $\theta_1, \theta_2 \neq \pi n/2$ ($n:$~integers), it is impossible that these gauge fields satisfy the periodic b.c. as long as there is only one flux tube. Therefore, we put two flux tubes in the transverse plane. In the following, we fix the angle parameters to be $\theta_1=0$ and $\theta_2 =\pi/4$, with which the initial electric and magnetic fields have the same field strength. For these angles, gauge configuration with two flux tubes located at $(x_1 ,y_1 )=(L_\perp/4 ,L_\perp/4 )$ and $(x_2 ,y_2 )=(3L_\perp/4 ,3L_\perp/4 )$ satisfies the periodic b.c. as long as $\Delta \ll L_\perp$. This configuration is realized by the functions
\begin{align}
\Theta_n (x,y) 
= \frac{\sqrt{\pi}}{2} Q_n \Delta &\bigg[ \text{Erf} \left( \frac{(x-x_1)\cos \theta_n +(y-y_1)\sin \theta_n }{\Delta} \right) \notag \\
&+\text{Erf} \left( \frac{(x-x_2)\cos \theta_n +(y-y_2)\sin \theta_n }{\Delta} \right) \bigg] \, .
\end{align}

Also the quark mode functions \eqref{qini_latt} must satisfy the periodic b.c. 
If the factors $V_n (\bxp ) = \exp \left[ i\Theta_n (\bxp ) \sigma^n /2 \right]$ are periodic, this requirement is fulfilled. 
For the angles $\theta_1=0$ and $\theta_2 =\pi/4$, the differences of $\Theta_n (x,y)$ at the boundaries are
\begin{align}
\Theta_1 (L_\perp ,y) -\Theta_1 (0,y) &= 2\sqrt{\pi} Q_1 \Delta \, , \\
\Theta_1 (x, L_\perp ) -\Theta_1 (x,0) &= 0 \, , \\
\Theta_2 (L_\perp ,y) -\Theta_1 (0,y) &= \sqrt{\pi} Q_2 \Delta \, , \\
\Theta_2 (x, L_\perp ) -\Theta_1 (x,0) &= \sqrt{\pi} Q_2 \Delta \, ,
\end{align}
for $\Delta \ll L_\perp$. Therefore, in order that the factors
\begin{equation}
V_n (\bxp ) = \cos \left[ \tfrac{1}{2} \Theta_n (\bxp )\right] +i\sigma^n \sin \left[ \tfrac{1}{2} \Theta_n (\bxp )\right]
\end{equation}
are periodic, the flux is quantized as
\begin{equation}
\sqrt{\pi} Q_1 \Delta = 2\pi n_1 \, , \hspace{10pt} 
\frac{1}{2} \sqrt{\pi} Q_2 \Delta = 2\pi n_2 \, , \hspace{10pt} (n_1 ,n_2 : \text{integers}).
\label{quantize}
\end{equation}

\section{Numerical results} \label{sec:result}
In this section, we present numerical results of solving the Yang--Mills equations (\ref{lattYM1}-\ref{lattYM4}) and the Dirac equation \eqref{Dirac} on the lattice for the axial charge production in the longitudinally expanding geometry.

\subsection{Uniform glasma} \label{subsec:result_uniform}
As a simple test for the real-time lattice computations of the axial charge production in the expanding geometry, we first consider the uniform glasma configuration introduced in Sec.~\ref{subsec:IC_uniform}. 
For background fields which are uniform not only in the $\eta$-direction but also in the transverse directions, the numerical cost to solve the Dirac equation is significantly reduced since the space-dependence of the mode functions is completely known.

We denote the typical energy scale of the glasma by $Q$, and initialize the gauge fields by setting $Q_1 = Q_2 = 2^{1/4} Q$, $\theta_1 = 0$, and $\theta_2 =\pi/4$.
This choice of the parameters results in the initial gauge fields
\begin{gather}
gA_1 (\tau =0) = 2^{1/4} Q \frac{\sigma^1}{2} +2^{-1/4} Q \frac{\sigma^2}{2} \, , \\
gA_2 (\tau =0) = 2^{-1/4} Q \frac{\sigma^2}{2} \, , \\
gA_\eta  (\tau =0) = 0 \, ,
\end{gather}
and 
\begin{equation}
gE_z  (\tau =0) = gB_z  (\tau =0) = Q^2 \frac{\sigma^3}{2} \, .
\end{equation}
Since all these fields are uniform, they trivially satisfy the periodic boundary condition.

As the initial condition for the quark mode functions, we employ the expression \eqref{qini_uni} after replacing the plane wave factor and the momenta by corresponding lattice expressions. 
By the replacement \eqref{latt_pw}, also the quark initial condition \eqref{qini_uni} satisfies the periodic boundary condition. 

In actual numerical computations, we cannot take the initial time $\tau_0$ to be exactly zero. Instead we take a small value of $\tau_0$ as $Q \tau_0 =10^{-3}$. We have confirmed that varying it between $5 \cdot 10^{-4}$ and $10^{-2}$ does not alter the later time behavior.
Unless otherwise noted, we use in this subsection the lattice parameters $N_\perp=48$, $N_\eta=512$, $Q L_\perp =20$, $L_\eta=60$, ($Qa_\perp =0.417$, $a_\eta=0.117$), and the Wilson parameters are fixed to $r_\perp =r_\eta=2$. 

\begin{figure}[t]
 \begin{center}
  \includegraphics[clip,width=7.2cm]{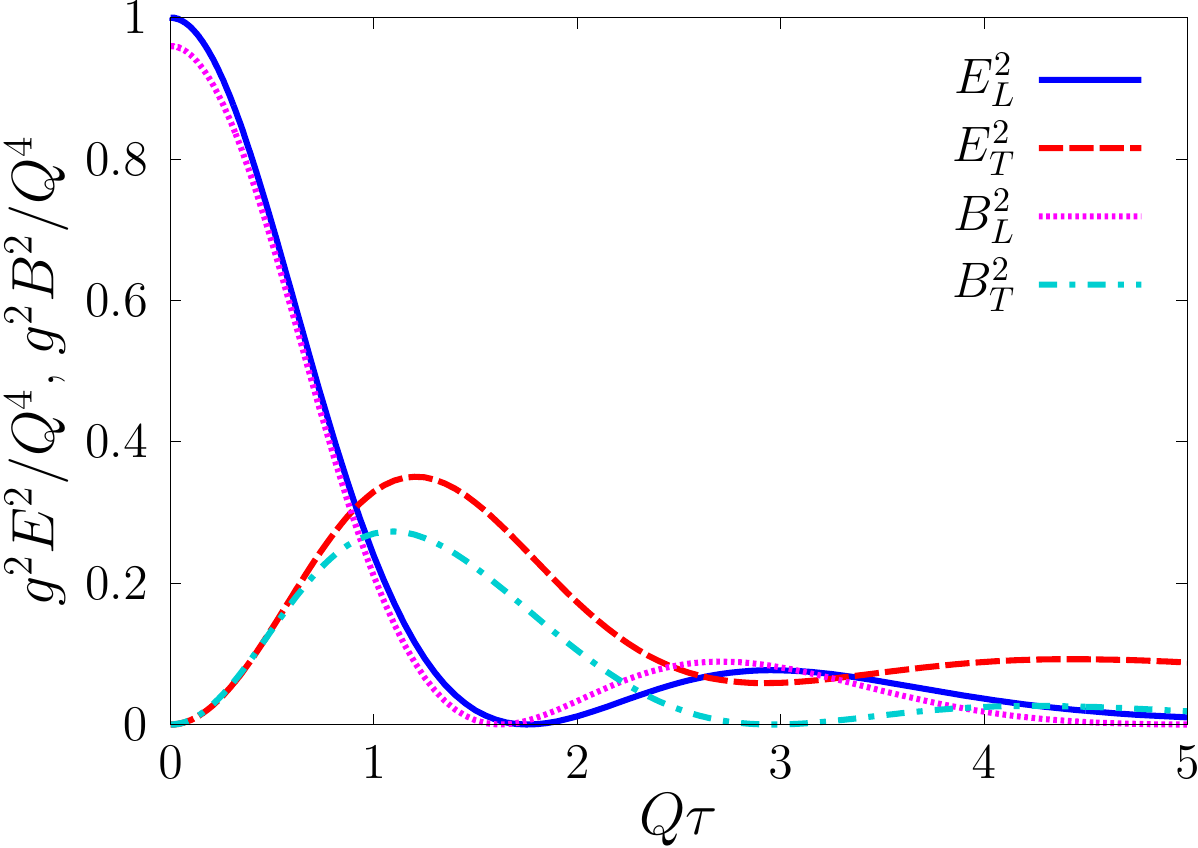} 
  \vspace{-8pt}
  \caption{The field strength of the uniform glasma as a function of the proper time. The longitudinal and the transverse components are plotted separately for the electric and the magnetic fields.}
  \label{fig:EandB_uni}
 \end{center}
\end{figure}
\begin{figure}
 \begin{center}
  \includegraphics[clip,width=7.2cm]{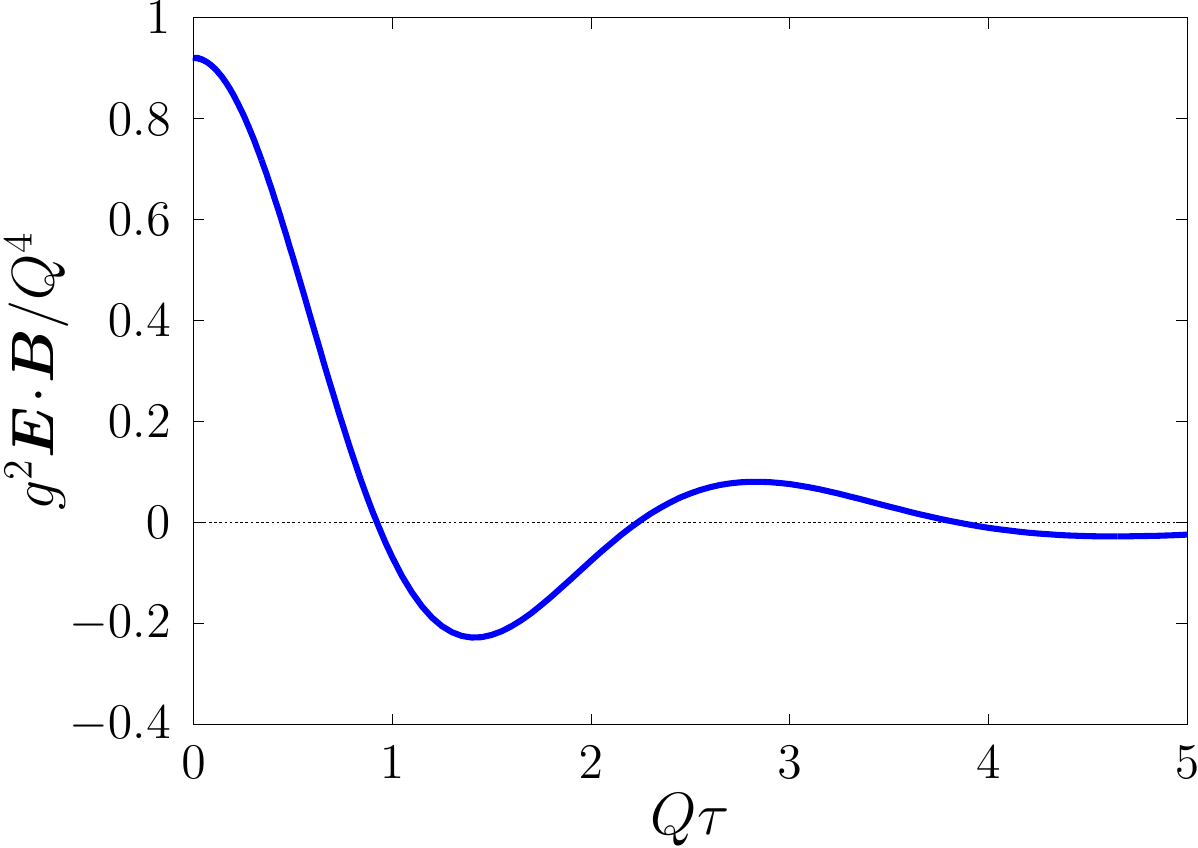} 
  \vspace{-8pt}
  \caption{Time evolution of $\bE^a \cd \bB^a$ for the uniform glasma.}
  \label{fig:EdotB_uni}
 \end{center}
\end{figure}

First, we show results of solving the Yang--Mills equations for the background gauge fields. The field strength of the longitudinal and transverse components is plotted for the electric and the magnetic fields separately as a function of time in Fig.~\ref{fig:EandB_uni}. The transverse components are defined as $E_T^2 = E_x^2 +E_y^2$ and $B_T^2 = B_x^2 +B_y^2$. 
In this and the following figures, all quantities are shown in dimensionless unit scaled by appropriate powers of $Q$.
Furthermore, the factor $g^2$ is multiplied to the field strength to make it order one.\footnote{%
In the leading order of the strong-field and weak-coupling approximation, the coupling $g$ appears in the equations only through the combination of $gA_\mu$. Therefore, we do not need to specify the value of the coupling in our computations.
}
The result shown in Fig.~\ref{fig:EandB_uni} looks similar to that with the MV model initial condition first shown in Ref.~\cite{Lappi:2006fp}; Initially only the longitudinal components are nonzero. As the longitudinal components decrease in time, the transverse components are induced, and eventually all the components decay in time. 
We point out that the decay of the fields is in fact nontrivial for the uniform system.
For example, if the initial field has only longitudinal electric component (which can be realized by e.g. $\theta_1 -\theta_2 = 0$), the field strength stays constant even in the longitudinally expanding system.\footnote{This situation is analogous to an electric field between two (infinitely large) capacitor plates, in which the field strength of the electric field is independent of the distance between the capacitor plates.}
This is because the nonlinear terms in the Yang--Mills equations do not play any role for that field configuration. 
Therefore, the decay of the uniform glasma seen in Fig.~\ref{fig:EandB_uni} is caused by the interplay between the system expansion and the nonlinear interaction of the color fields. 

\begin{figure}[t]
 \begin{center}
  \includegraphics[clip,width=7.5cm]{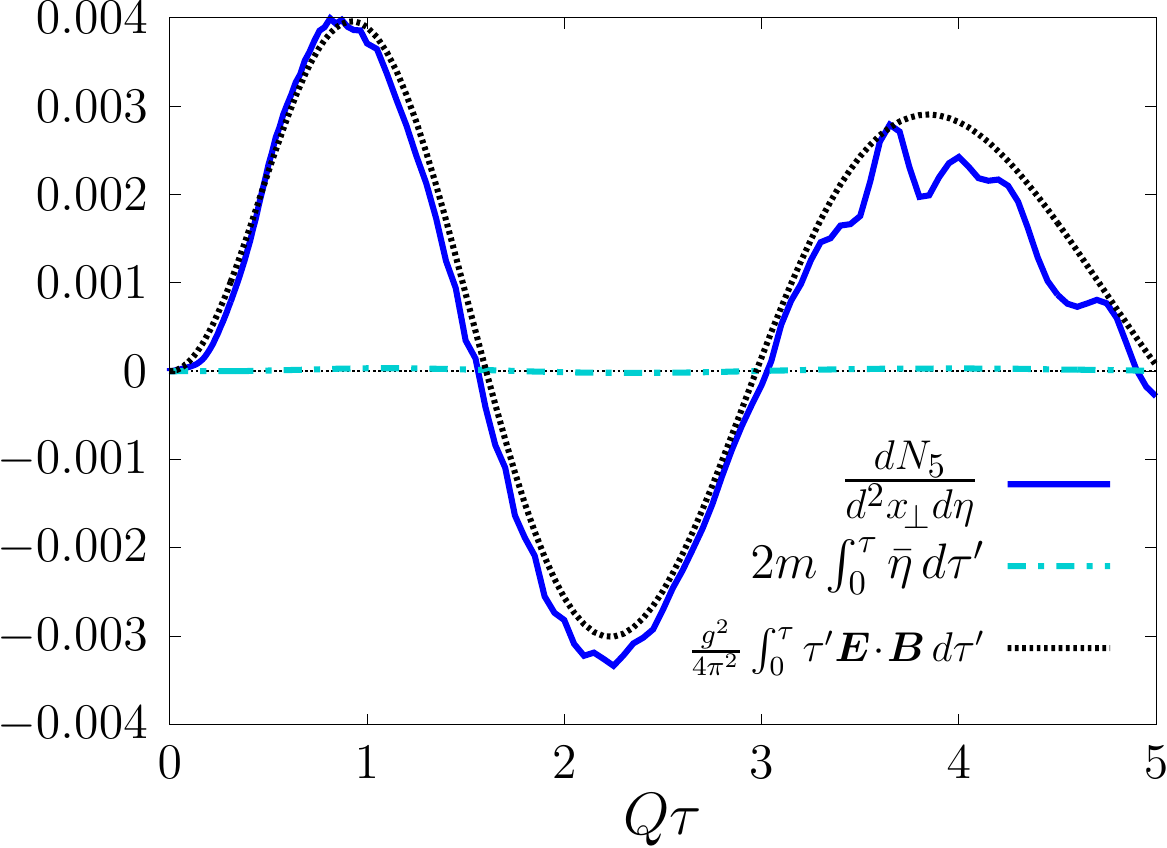} 
  \vspace{-8pt}
  \caption{The terms in the anomaly relation \eqref{budget_uni} are plotted separately as a function of time for quark mass $m/Q=0.01$. All the quantities are made dimensionless being divided by the factor $Q^2$.}
  \label{fig:ano_uni}
 \end{center}
\end{figure}

As shown in Fig.~\ref{fig:EdotB_uni}, $\bE^a \cd \bB^a$ exhibits damped oscillation in time. Due to the oscillation, $\bE^a \cd \bB^a$ changes its sign leading to nonmonotonic behavior of the axial charge density as we will discuss below.

In a uniform system, the transverse divergence term of the axial current disappears, and the anomaly relation leads
\begin{equation}
\frac{dN_5}{d^2 x_{\!\perp} d\eta} 
= 2m \int_0^\tau \! \bar{\eta} (\tau^\pr ) d\tau^\pr
+\frac{g^2}{4\pi^2} \int_0^\tau \! \tau^\pr \bE^a \cd \bB^a d\tau^\pr \, .
\label{budget_uni}
\end{equation}
Each term in this equation is plotted as a function of time in Fig.~\ref{fig:ano_uni} for quark mass $m/Q=0.01$. 
For such light quark mass, the pseudoscalar condensate term $2m \int_0^\tau \!d\tau^\pr \, \bar{\eta}$ is negligible, and hence the remaining two terms must agree for the realization of the axial anomaly on the lattice. Indeed, the axial charge density and the time-integration of the topological charge density show an agreement, though we see some deviations especially at later time. This result demonstrates that the axial anomaly can be described by the Wilson fermion even in the longitudinally expanding geometry.

\begin{figure}[t]
 \begin{center}
  \includegraphics[clip,width=7.5cm]{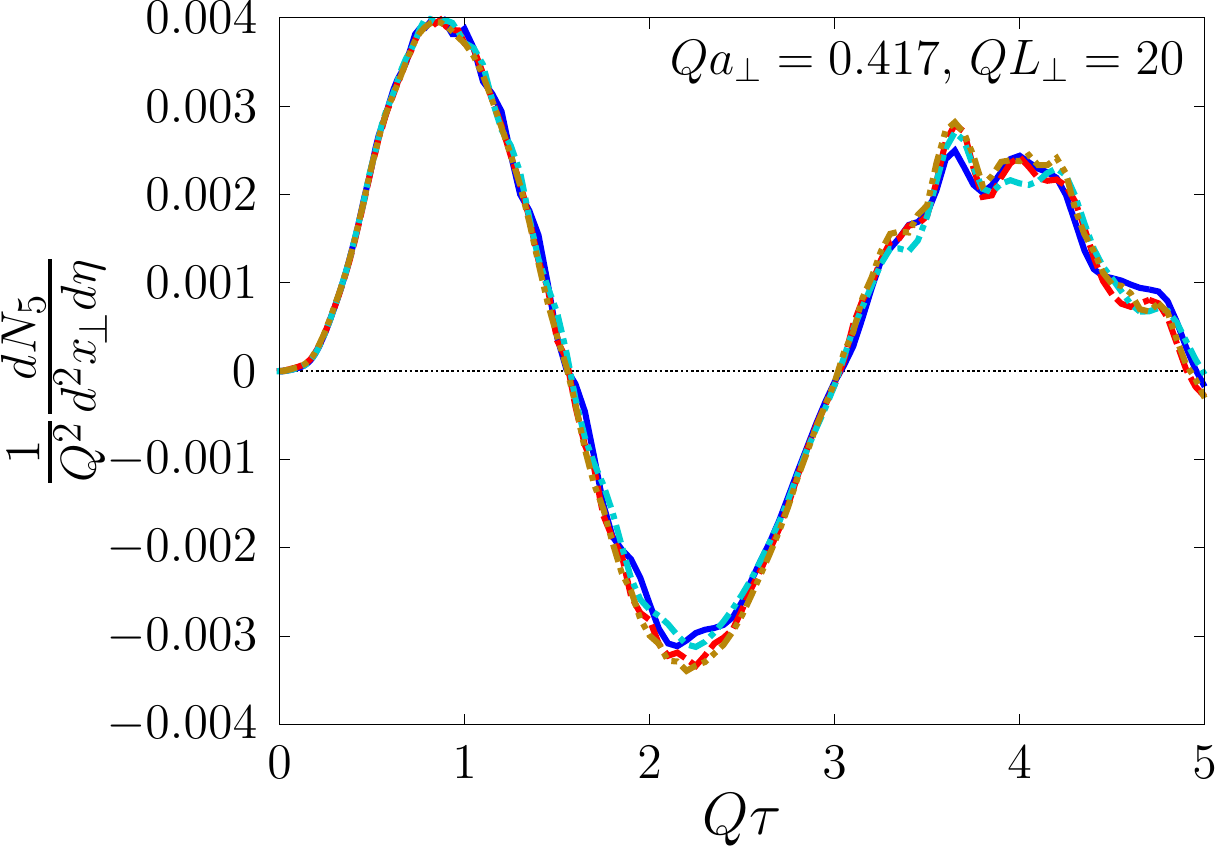} 
  \includegraphics[clip,width=4cm]{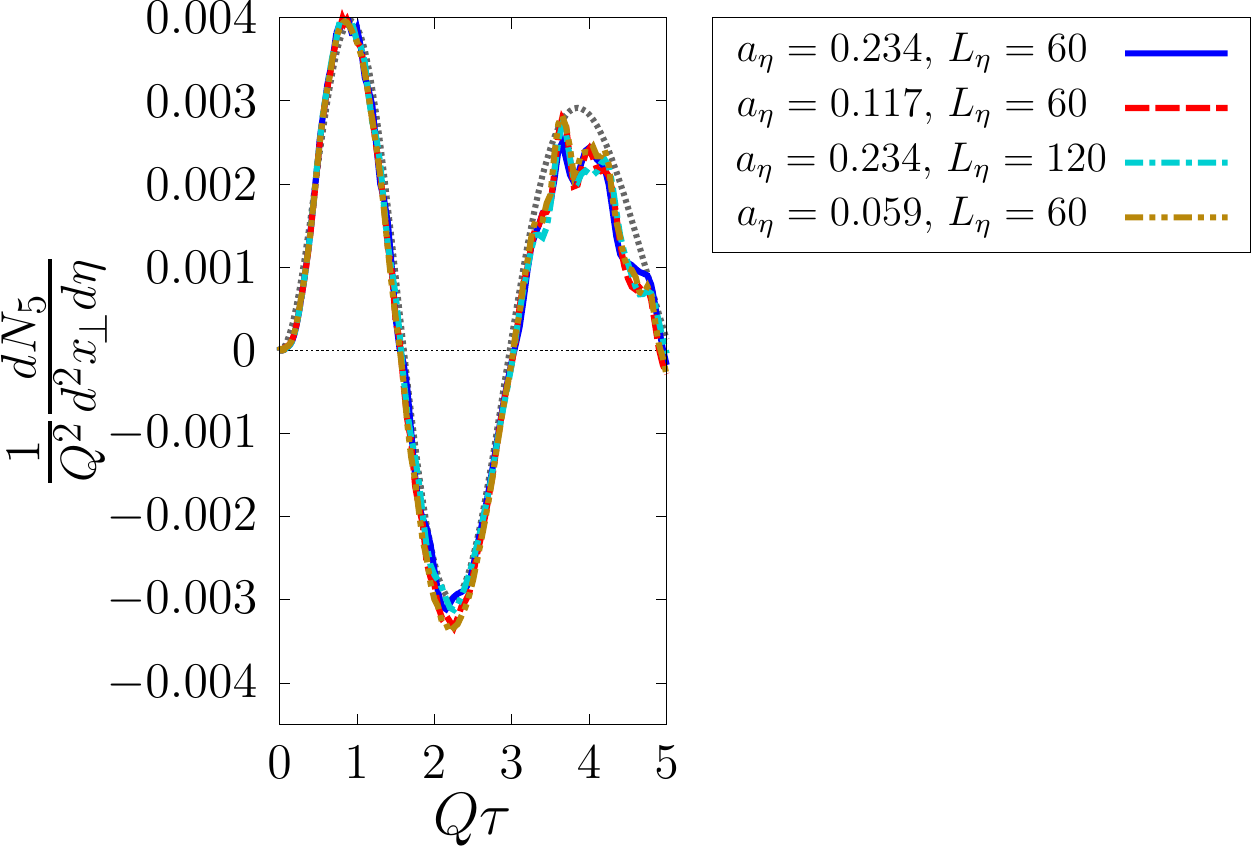}
  \vspace{-8pt}
  \caption{The axial charge density for different longitudinal lattice parameters. The quark mass is $m/Q=0.01$. The transverse lattice parameters are fixed to $Qa_\perp =0.417$ and $QL_\perp=20$.}
  \label{fig:ls_l}
 \end{center}
\end{figure}

\begin{figure}
 \begin{center}
  \includegraphics[clip,width=7.5cm]{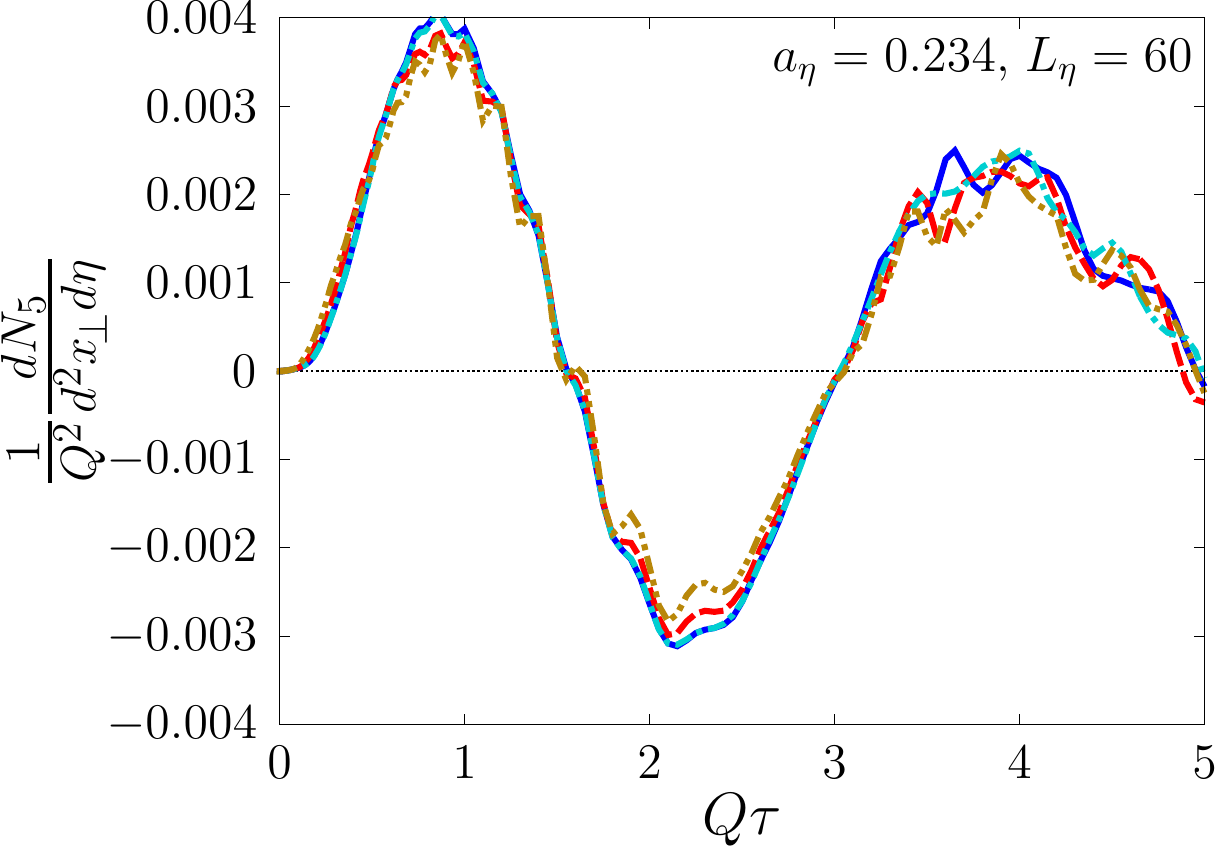} 
  \includegraphics[clip,width=4.7cm]{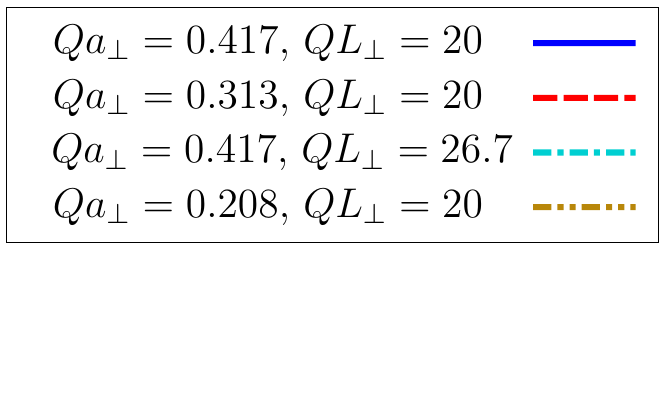}
  \vspace{-8pt}
  \caption{The axial charge density for different transverse lattice parameters. The quark mass is $m/Q=0.01$. The longitudinal lattice parameters are fixed to $a_\eta =0.234$ and $L_\eta=60$.}
  \label{fig:ls_t}
 \end{center}
\end{figure}

In Figs.~\ref{fig:ls_l} and \ref{fig:ls_t}, numerical results for the axial charge density computed with different lattice parameters are shown. Both for the longitudinal (Fig.~\ref{fig:ls_l}) and the transverse (Fig.~\ref{fig:ls_t}) lattice parameters, the results are nearly insensitive to the changes of either the UV cutoff scale $1/a$ and the infrared scale $1/L$. 


\begin{figure}[t]
 \begin{center}
  \includegraphics[clip,width=7.5cm]{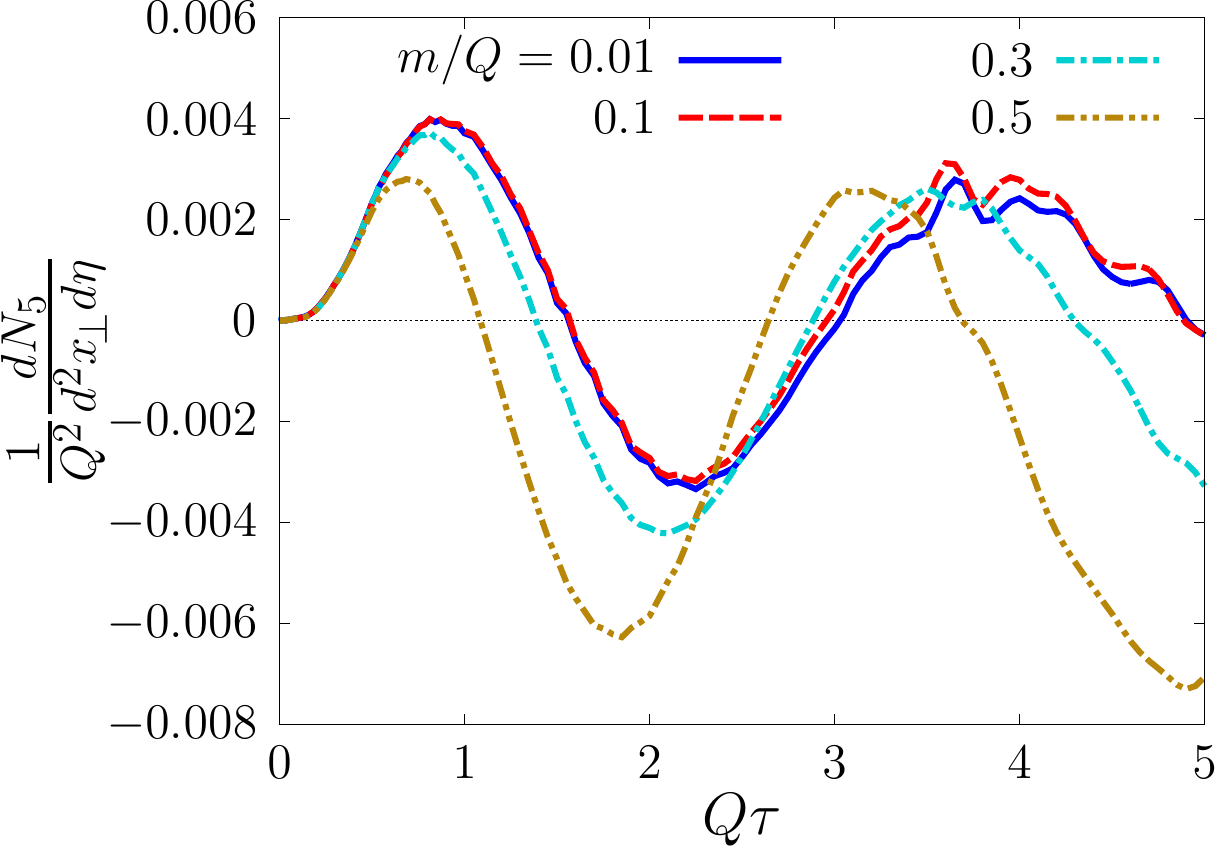} 
  \vspace{-8pt}
  \caption{Time evolution of the axial charge density for different quark masses.}
  \label{fig:mass_dep}
 \end{center}
\end{figure}

\begin{figure}
 \begin{center}
  \includegraphics[clip,width=7.5cm]{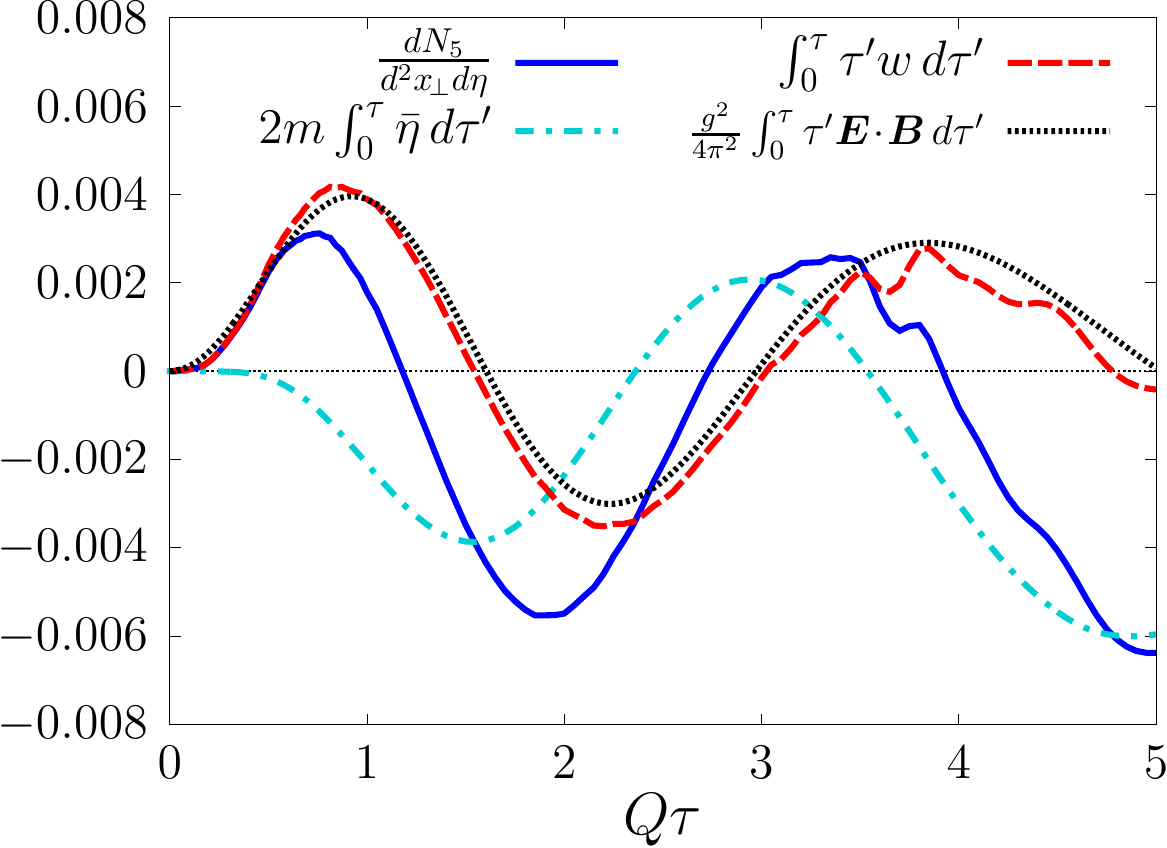} 
  \vspace{-8pt}
  \caption{The terms in the anomaly relation \eqref{budget_uni} are plotted separately as a function of time for quark mass $m/Q=0.5$. Also the Wilson term contribution $\int_0^\tau \tau^\prime w\, d\tau^\prime$ is plotted for comparison.}
  \label{fig:m0.5}
 \end{center}
\end{figure}

So far, we have shown numerical results for light quark mass $m/Q=0.01$, in which case the pseudoscalar condensate term is negligible. We now present in Fig.~\ref{fig:mass_dep} the dependence of the axial charge density on the quark mass.
In the computations with the quark masses $m/Q=0.3$ and 0.5, we have used the technique of the Wilson parameter averaging \cite{Mace:2016shq}. 
For lighter masses, $m/Q=0.01$ and 0.1, the curves are nearly overlapped indicating these quarks can be regarded as almost massless. In contrast, the results for heavier quarks, $m/Q=0.3$ and 0.5, show significant deviations from those for the light quarks. This is because the pseudoscalar condensate term is comparable to other terms in the anomaly relation \eqref{budget_uni} for these masses. 
To illustrate it, we plot the terms in the anomaly relation \eqref{budget_uni} separately for quark mass $m/Q=0.5$ in Fig.~\ref{fig:m0.5}. Also the time-integral of the Wilson term contribution \eqref{Wilson_cont} is depicted. 
The agreement between the Wilson term contribution $\int_0^\tau \tau^\prime w\, d\tau^\prime$ and the gauge field contribution $\tfrac{g^2}{4\pi^2} \int_0^\tau \tau^\prime \bE^a \cd \bB^a \, d\tau^\prime$ indicates the realization of the axial anomaly. 
For this quark mass, the pseudoscalar condensate term is indeed as large as other terms especially at later times. 
However, at very early times, $Q\tau \ltsim 0.5$, the rise of the pseudoscalar condensate term is slower than the other terms. This is the reason why the axial charge densities show little dependence on the quark masses at the early times in Fig.~\ref{fig:mass_dep}. 
At the later times, both of the axial charge density and the pseudoscalar condensate term show oscillation.
Interestingly, their oscillation phases are different, and therefore the pseudoscalar condensate does not always diminish the axial charge density in this oscillating background field.


Before closing this subsection, we show a rough estimate of the axial charge density in physical units for the numerical results obtained in this subsection. 
All the quantities so far have been plotted in the dimensionless unit scaled by the typical energy scale of the glasma $Q$.  
In this dimensionless unit, the axial charge density $\frac{1}{Q^2} \frac{dN_5}{d^2 x_{\!\perp} d\eta} \approx 0.004$ is produced at the time $Q\tau=1$ for nearly massless quarks.
In relativistic heavy-ion collisions at RHIC and the LHC, the typical energy scale of the glasma is order of 1 GeV.
For $Q=1$ GeV, the value $\frac{1}{Q^2} \frac{dN_5}{d^2 x_{\!\perp} d\eta} \approx 0.004$ is translated to $\frac{dN_5}{d^2 x_{\!\perp} d\eta} \approx 0.1$/fm$^2$, which means 0.1 excess of right-handed quarks over left-handed quarks per flavor in a unit volume with transverse area of 1 fm$^2$ and unit space-time rapidity.
Of course, these values should not be taken too seriously because the color field configuration considered in the calculation is not so realistic; it is uniform in the transverse plane and the color directions of the fields are chosen such that $\mathbf{E}^a \cd \mathbf{B}^a$ is maximum for fixed energy density. As we see in the next subsection, the amount of axial charge density is reduced for inhomogeneous configurations because the spatial divergence term of the axial current takes some fraction in the anomaly relation. Also considering the color SU(3) theory instead of SU(2) employed in study would modify the results quantitatively. 

\subsection{Glasma flux tubes} \label{subsec:result_flux}
In the previous subsection, we have confirmed that the axial anomaly on the real-time lattice in the expanding geometry can be described by using the Wilson fermion for the uniform glasma configuration. In this subsection, we present numerical results for the flux tube configuration, which is inhomogeneous in the transverse plane. 
The results presented in this subsection are computed with the lattice parameters $N_\perp=64$, $N_\eta=256$, $Q L_\perp =30$, $L_\eta=60$, ($Qa_\perp =0.469$, $a_\eta=0.234$), the Wilson parameters $r_\perp =r_\eta=2$, and the quark mass $m/Q=0.01$. 

\begin{figure}[t]
 \begin{tabular}{cc}
 \begin{minipage}{0.5\hsize}
  \begin{center}
   \includegraphics[clip,width=6cm]{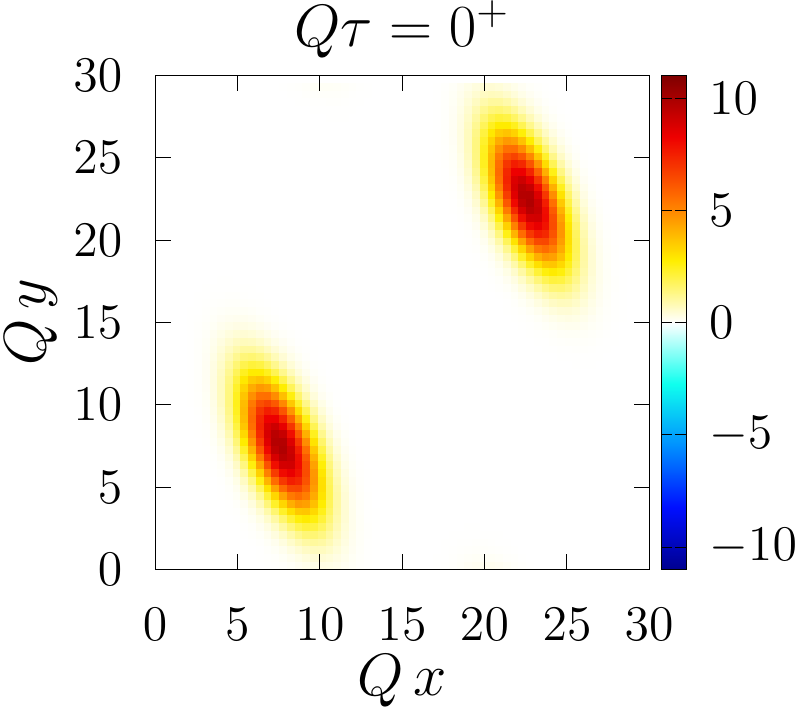}
  \end{center}
 \end{minipage} &
 \begin{minipage}{0.5\hsize}
  \begin{center}
   \includegraphics[clip,width=6cm]{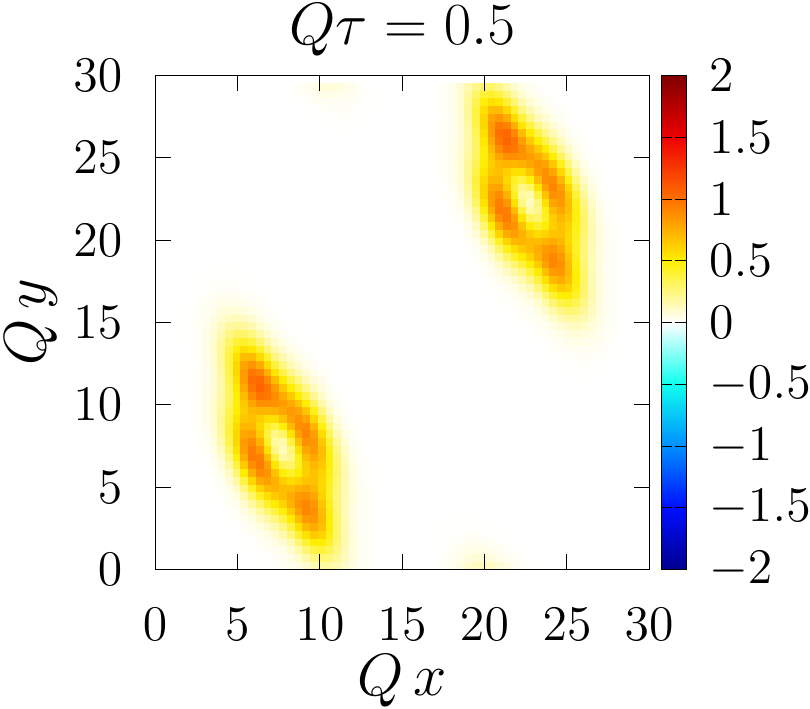}
  \end{center}
 \end{minipage} \\ \ & \ \\
 \begin{minipage}{0.5\hsize}
  \begin{center}
   \includegraphics[clip,width=6cm]{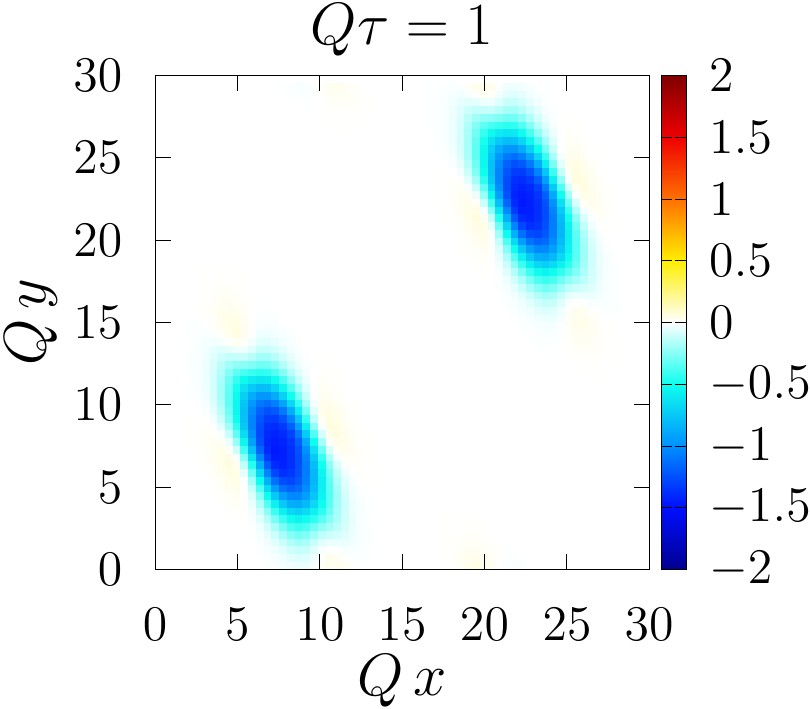}
  \end{center}
 \end{minipage} &
 \begin{minipage}{0.5\hsize}
  \begin{center}
   \includegraphics[clip,width=6cm]{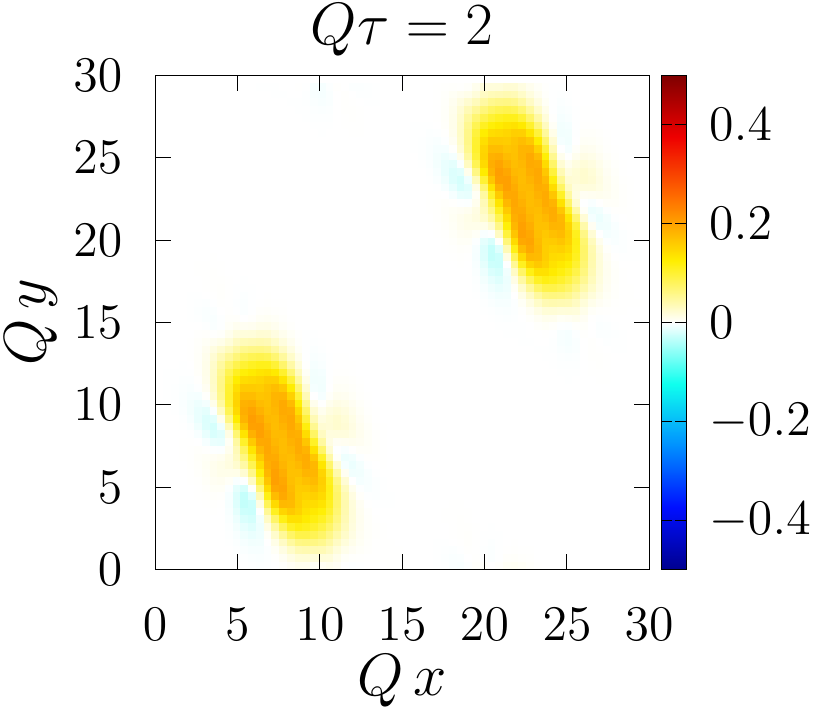}
  \end{center}
 \end{minipage} \\ \ & \ \\
 \begin{minipage}{0.5\hsize}
  \begin{center}
   \includegraphics[clip,width=6cm]{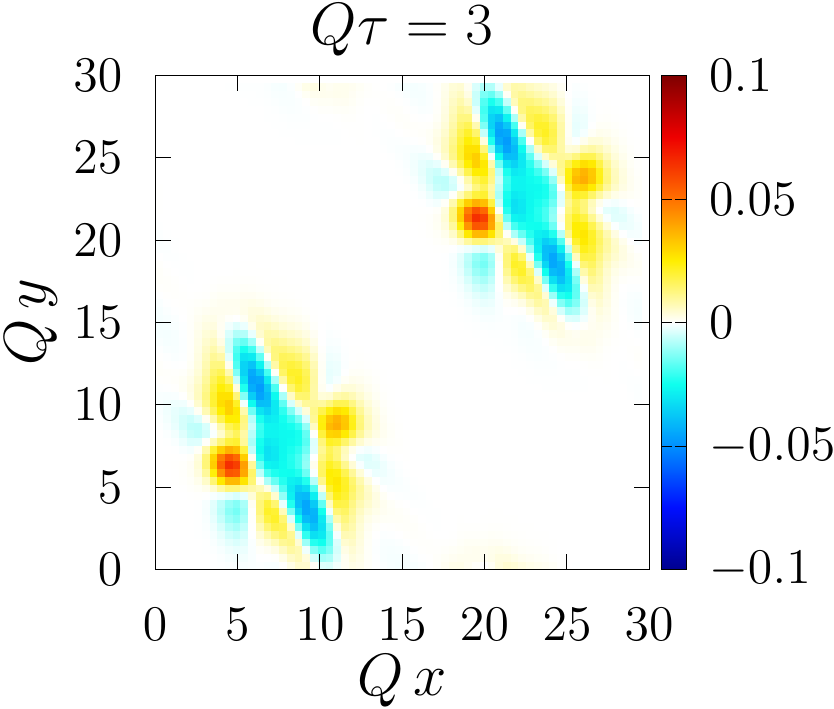}
  \end{center}
 \end{minipage} &
 \begin{minipage}{0.5\hsize}
  \begin{center}
   \includegraphics[clip,width=6cm]{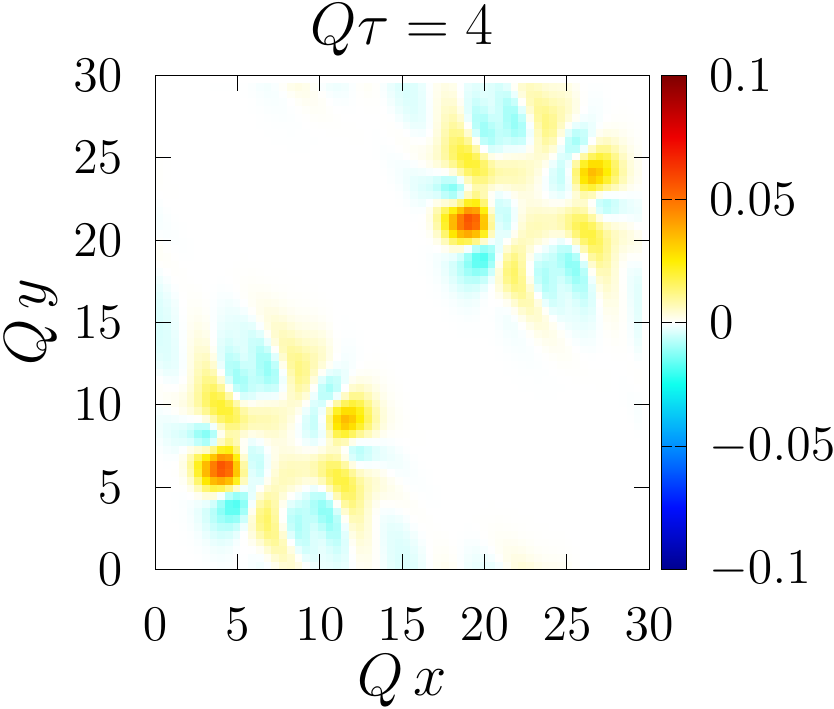}
  \end{center}
 \end{minipage} 
 \end{tabular}
\caption{Density plots of $g^2 \bE^a \cd \bB^a /Q^4$ in the transverse plane for the glasma flux tube configuration at the initial time (upper left) and later times $Q\tau=0.5$, 1, 2, 3, 4.}
\label{fig:EdotB2d}
\end{figure}

As discussed in Sec.~\ref{subsec:bc}, we need to put two flux tubes and their flux is quantized as \eqref{quantize} for specific values of the angle parameters $\theta_1 =0$ and $\theta_2=\pi/4$ due to the requirement of the periodic b.c. on the transverse lattice.
In the time range $Q\tau \leq 5$ that we consider in the following, the two flux tubes are causally separated and thus our numerical simulations are essentially equivalent to those of one flux tube. 
We choose parameters $Q_1 =Q_2 = 4\sqrt{\pi}Q/3$ and $\Delta =3/Q$.
The initial transverse profile of $\bE^a \cd \bB^a$ is depicted in the upper left panel of Fig.~\ref{fig:EdotB2d} as a density plot. It has distorted Gaussian shapes.

\begin{figure}[tb]
 \begin{center}
  \includegraphics[clip,width=7.5cm]{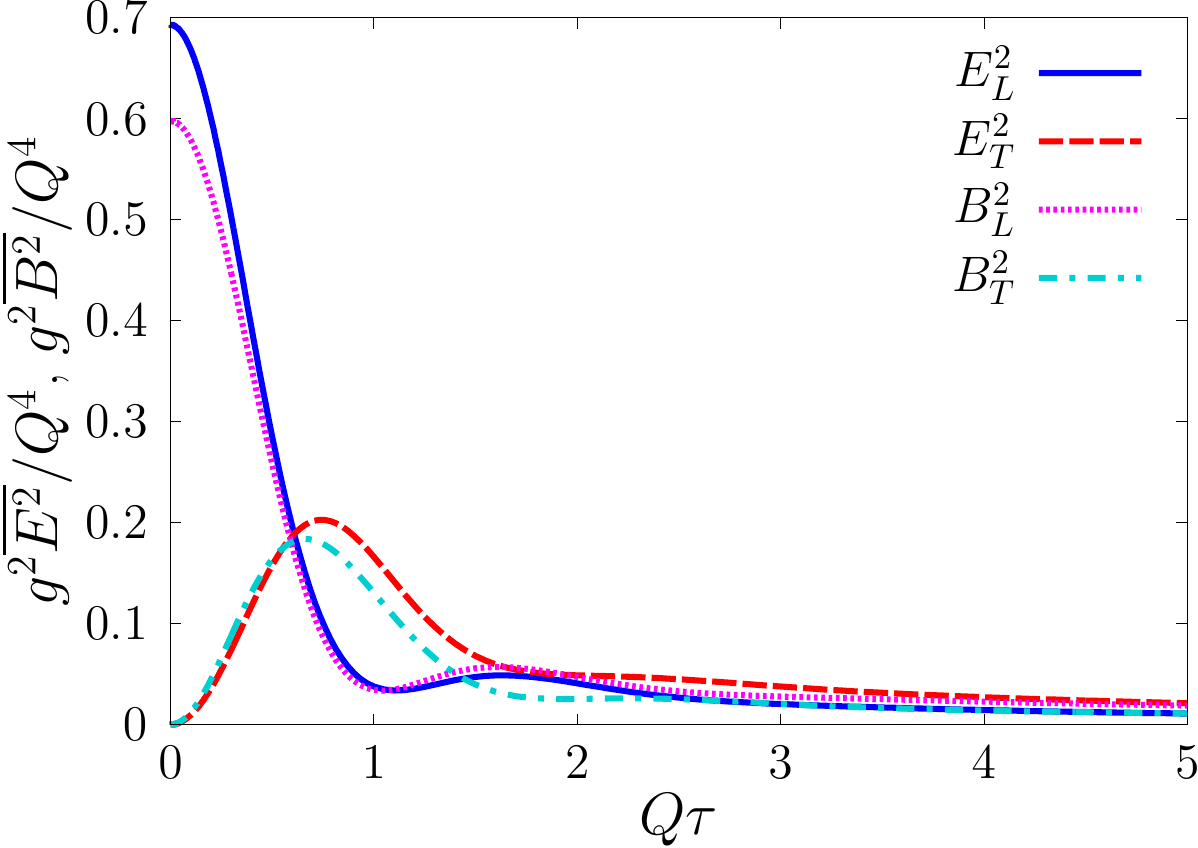} 
  \vspace{-8pt}
  \caption{The space-averaged field strength for the glasma flux tube configuration as a function of the proper time. The longitudinal and the transverse components are plotted separately for the electric and the magnetic fields.}
  \label{fig:EandB_ft}
 \end{center}
\end{figure}

\begin{figure}[tb]
 \begin{center}
  \includegraphics[clip,width=7.5cm]{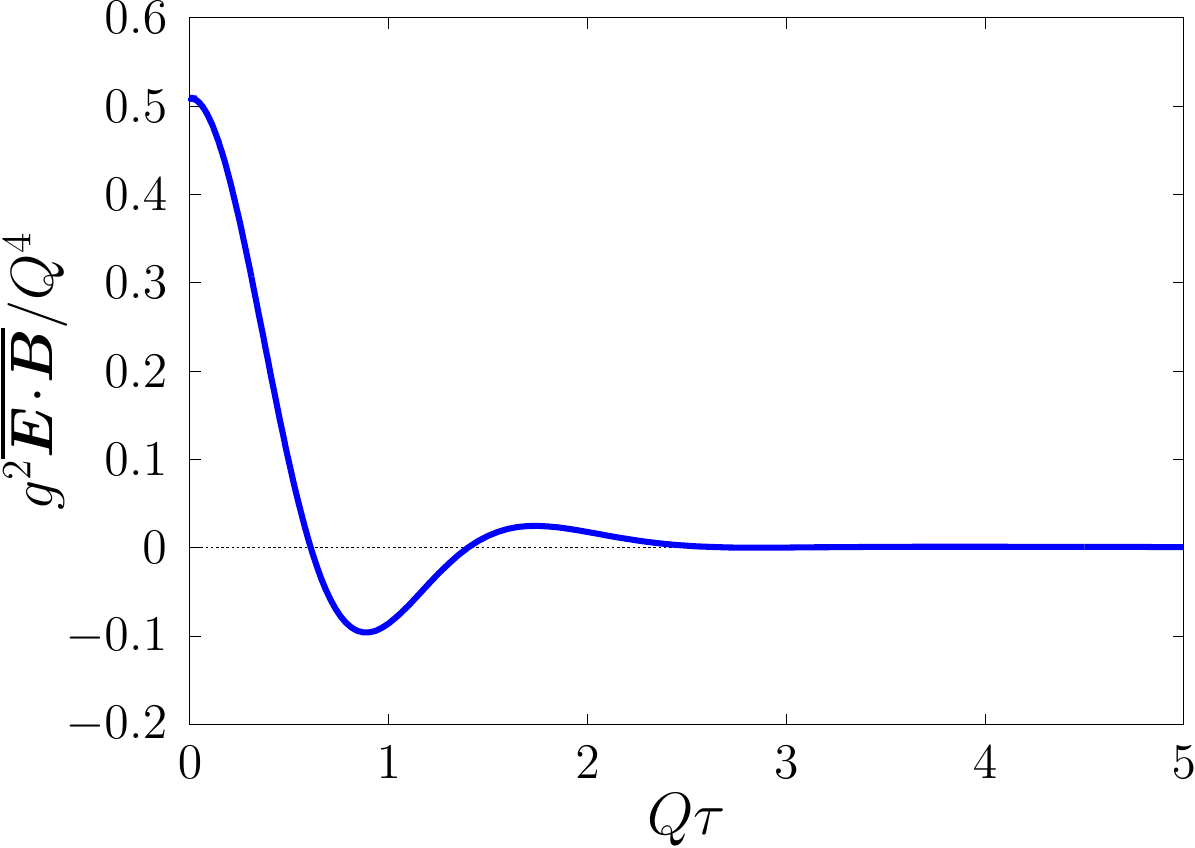} 
  \vspace{-8pt}
  \caption{The space-averaged $\bE^a \cd \bB^a$ for the glasma flux tube configuration as a function of the proper time. }
  \label{fig:EdotB_ft}
 \end{center}
\end{figure}

Other panels in Fig.~\ref{fig:EdotB2d} exhibit the profiles of $\bE^a \cd \bB^a$ at later times after evolved by the Yang--Mills equations. 
At earlier times than the time scale characterized by the flux tube width $\Delta =3/Q$, the propagation of the fields in the transverse plane is not obvious. In this early time stage, the fields possess strong coherence and show oscillation like the uniform field. At later times, $Q\tau=3$ and 4, the propagation of the fields in the transverse plane becomes more apparent, and the fields lose coherence.
These observations are reinforced by Figs.~\ref{fig:EandB_ft} and \ref{fig:EdotB_ft}, where space-averaged field strength and $\bE^a \cd \bB^a$, respectively, are plotted as a function of time. 
In the following, the space averaging is denoted by an overline like $\overline{E^2}$. 
At the earlier time stage, the space-averaged $\bE^a \cd \bB^a$ shows damped oscillation similar to that of the uniform glasma shown in Fig.~\ref{fig:EdotB_uni}. 
At the later time stage, the decay of $\bE^a \cd \bB^a$ is much faster than the uniform case because of the decoherence of the fields due to the transverse propagation.
We note that the fields remain coherent in the longitudinal direction since the boost invariance is strictly maintained in our computations. The longitudinal coherence can be broken by instabilities if we introduce rapidity-dependent fluctuations in the initial condition for the gauge fields \cite{Romatschke:2005pm,Fukushima:2011nq}. 

\begin{figure}[tb]
 \begin{center}
  \includegraphics[clip,width=7.5cm]{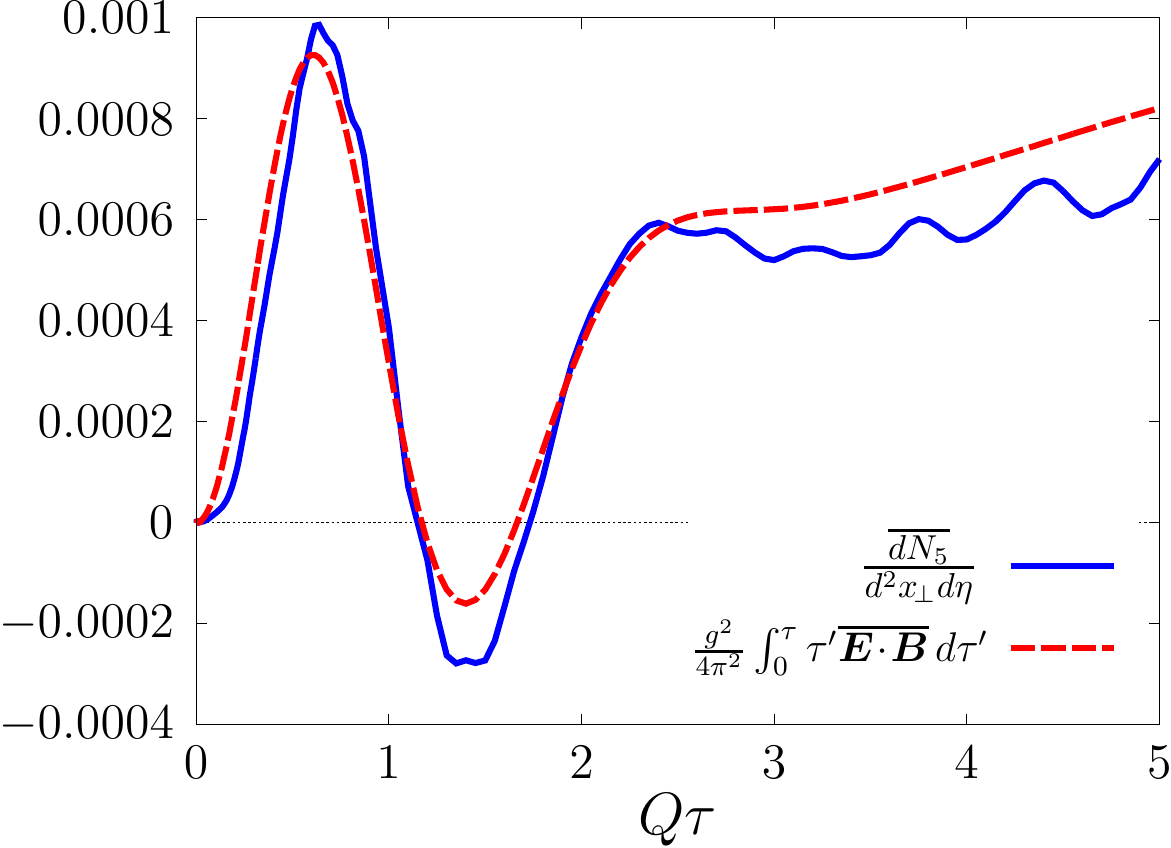} 
  \vspace{-8pt}
  \caption{The space-averaged axial charge density compared to the $\bE^a \cd \bB^a$ contribution in the space-averaged anomaly relation \eqref{av_anom}.}
  \label{fig:av_anom}
 \end{center}
\end{figure}

Next, we present numerical results of solving the Dirac equation under the glasma flux tube configuration for the axial charge production. First, let us look at the anomaly relation averaged over space,
\begin{equation}
\overline{\frac{dN_5}{d^2 x_{\!\perp} d\eta} }
= \frac{g^2}{4\pi^2} \int_0^\tau \! \tau^\pr \overline{\bE^a \cd \bB^a} \, d\tau^\pr \, ,
\label{av_anom}
\end{equation}
where the pseudoscalar condensate term is dropped out since it is negligible for $m/Q=0.01$, and the transverse divergence term of the axial current is absent as it disappears by the space averaging. 
This relation is examined in Fig.~\ref{fig:av_anom}, where the two terms in both sides of the equation are plotted separately as a function of time. Although there are noticeable deviations, the overall behaviors of the two curves roughly agree, demonstrating the realization of the axial anomaly in this transversally inhomogeneous system. 
Compared to the corresponding result in the uniform system shown in Fig.~\ref{fig:ano_uni}, what is remarkable is the nearly monotonic increase at later times $Q\tau \gtsim 3$. In the uniform glasma, $\bE^a \cd \bB^a$ continues to oscillate even at the later times, and hence also its time integral does. In the flux tube configuration, $\bE^a \cd \bB^a$ decays faster, and this decaying behavior rather helps nonzero axial charge density remain at the later times. For example, if $\bE^a \cd \bB^a$ decays as $1/\tau$, the time-integral $\int_0^\tau \tau^\prime \bE^a \cd \bB^a \, d\tau^\prime$ increases linearly in time. A similar observation, axial charges persist to be present after coherent gauge fields die out due to an  instability, has been made in Ref.~\cite{Tanji:2016dka} for a nonexpanding system.

\begin{figure}[tb]
 \begin{tabular}{cc}
 \begin{minipage}{0.5\hsize}
  \begin{center}
   \includegraphics[clip,width=7.5cm]{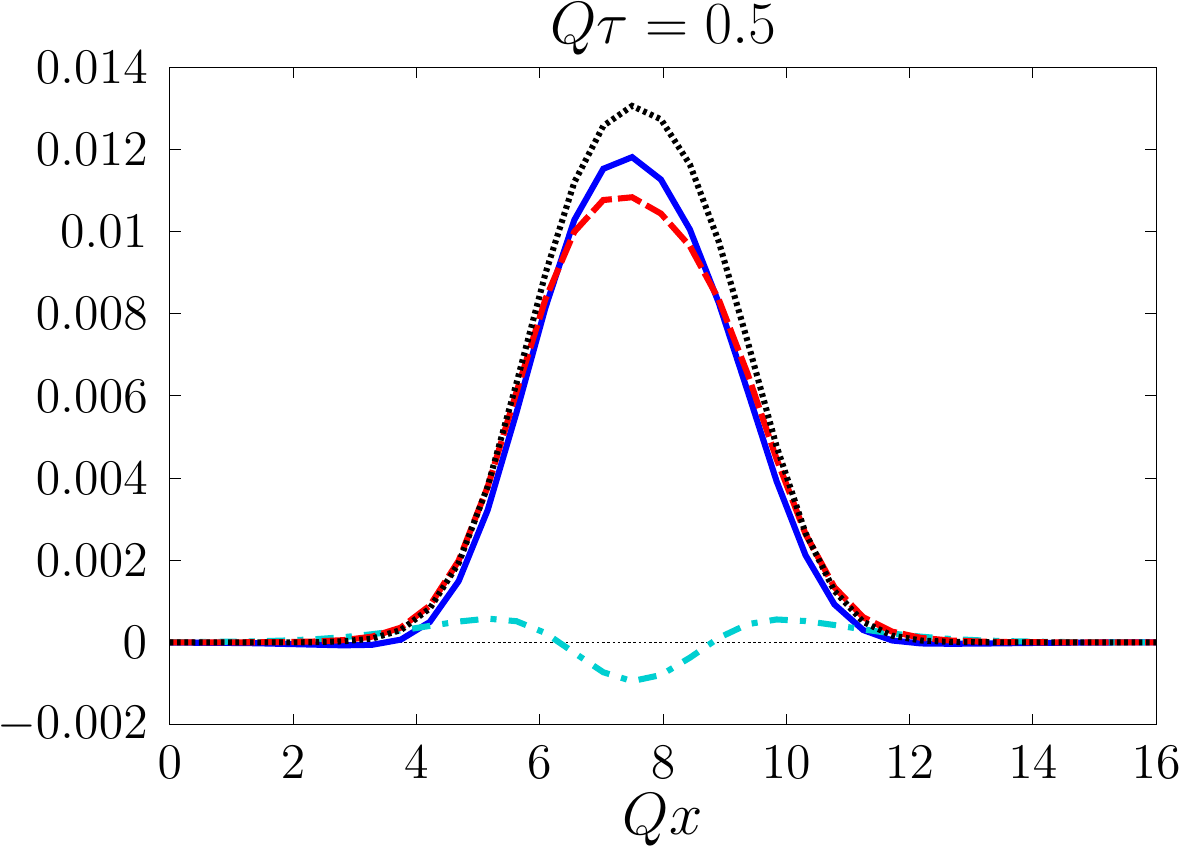}
  \end{center}
 \end{minipage} &
 \begin{minipage}{0.5\hsize}
  \begin{center}
   \includegraphics[clip,width=7.5cm]{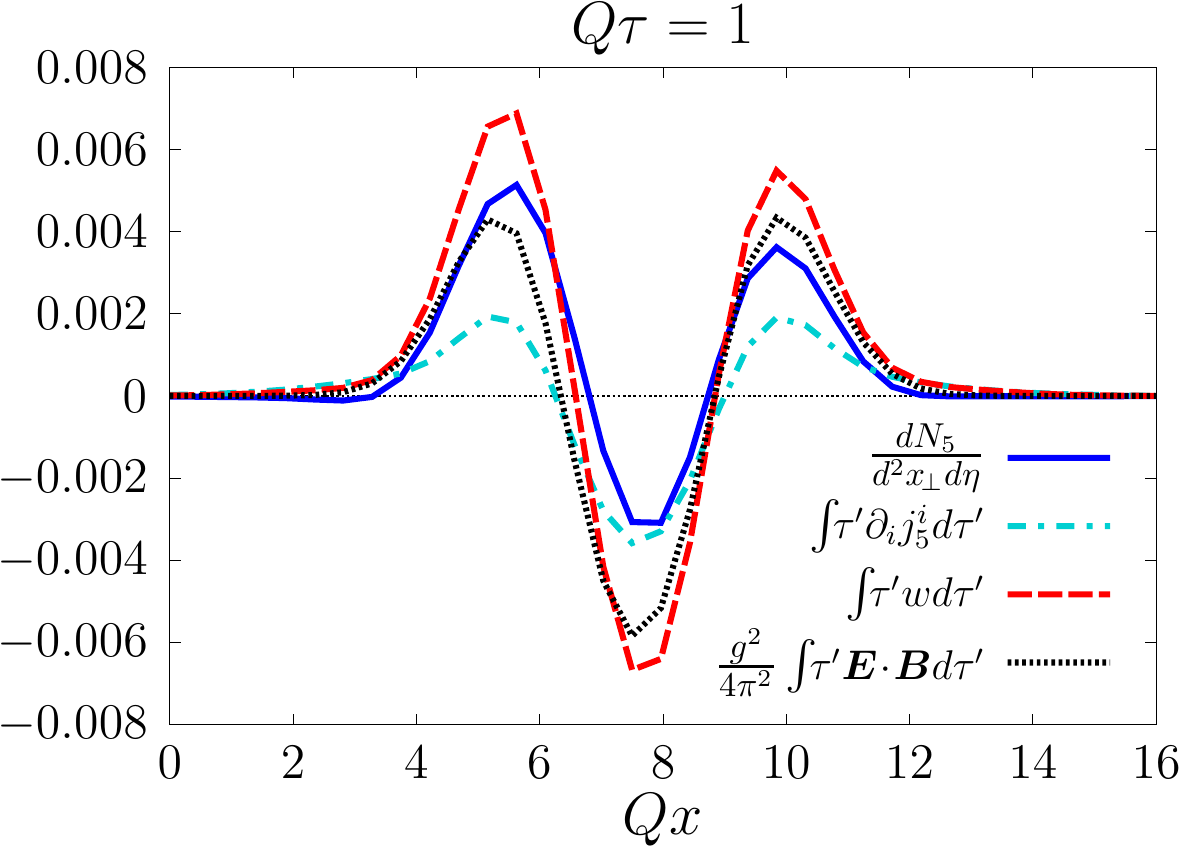}
  \end{center}
 \end{minipage} 
 \end{tabular}
\caption{The terms in the anomaly relation \eqref{budget} are plotted separately as a function of the transverse coordinate $x$ for fixed $y$-coordinate, $y=L_\perp/4$. Left: at $Q\tau=0.5$. Right: at $Q\tau=1$. 
Also the Wilson term contribution $\int_0^\tau \tau^\pr w\, d\tau^\pr$ is plotted for comparison.}
\label{fig:comp_ft}
\end{figure}

Lastly, we examine the anomaly relation \eqref{budget} as a function of the space coordinate.
In Fig.~\ref{fig:comp_ft}, all the terms in Eq.~\eqref{budget} are plotted separately as a function of the coordinate $x$ for fixed $y=L_\perp/4$ at times $Q\tau=0.5$ and $Q\tau=1$. Also the Wilson term contribution $\int_0^\tau \tau^\pr w\, d\tau^\pr$ is plotted for comparison. 
The pseudoscalar condensate term is not depicted since it is negligible. 
Now, the relation
\begin{equation}
\frac{dN_5}{d^2 x_{\!\perp} d\eta} +\int_0^\tau \! \tau^\pr \partial_i j_5^i \, d\tau^\pr
= 2m \int_0^\tau \! \bar{\eta} \, d\tau^\pr +\int_0^\tau \! \tau^\pr w \, d\tau^\pr 
\end{equation}
is trivially satisfied as we solve the lattice Dirac equation \eqref{Dirac}. 
Therefore, the agreement between the terms $\int_0^\tau \tau^\prime \bE^a \cd \bB^a \, d\tau^\prime$ and $\int_0^\tau \tau^\pr w\, d\tau^\pr$ is the condition for the realization of the anomaly relation \eqref{budget}. 
Both at $Q\tau =0.5$ and 1, the overall behavior of the curves for these two terms roughly agree although at the most 20--40\% of deviations are present. 
We infer that these local deviations are due to insufficient resolutions at UV scales. 
In an inhomogeneous system, resolutions at small scales in the coordinates space, or equivalently UV scales in the momentum space, is more important than in a uniform system. 
Since the Wilson term changes the UV sector of the theory, space-dependent quantities may be more affected by it. In our system, the situation is further complicated by the fact that the longitudinal momentum scales vary rapidly in time as $1/\tau$. 
We expect that the use of finer lattices would improve the accuracy. However, it is extraordinary challenging due to expensive numerical cost for the mode function method. Improvement of the computational method is desirable. 
We leave these issues for future investigations.

Having compromised with this accuracy of the anomaly relation in the present study, let us compare the axial charge density $\frac{dN_5}{d^2x_{\!\perp}d\eta}$ and the the spatial divergence term of the axial current $\int_0^\tau \tau^\pr \partial_i j_5^i \, d\tau^\pr$. 
The latter term characterizes the outflow of axial charge. 
At the very early time, $Q\tau=0.5$, the outflow term $\int_0^\tau \tau^\pr \partial_i j_5^i \, d\tau^\pr$ is much smaller than other terms. In this case, one can compute the axial charge density directly from $\bE^a \cd \bB^a$ without solving the Dirac equation similarly to the uniform system. Already at $Q\tau=1$, however, the outflow term becomes comparable to other terms signaling the propagation of the axial charge in the transverse plane.  
When the outflow term is not negligible, one cannot predict the amount of axial charge anymore directly from the anomaly relation. It is necessary to compute the full quantum dynamics of the quark fields by solving the Dirac equation as we have done in this study.

\section{Conclusions} \label{sec:conclusion}
In this paper we have investigated the axial charge production in the early stage of heavy-ion collisions by using the real-time lattice simulation method for classical gauge fields and quantum quark fields. 
To consistently include the effects of the colliding nuclei on the evolution of the quark fields, the solution of the Dirac equation under the CGC gauge fields has been used for the initial condition of the numerical computation for the time evolution after the collision. 

First, we considered the uniform configuration for the glasma gauge fields, and demonstrated that the Wilson fermion method generalized to the expanding geometry can correctly describe the axial anomaly on the lattice. In case of the uniform configuration, the color electromagnetic fields continue to coherently oscillate.
Consequently, the axial charge density per unit rapidity exhibits oscillating behavior.
Next, we computed the evolution of the glasma flux tubes. Due to the longitudinal expansion and the dynamics in the transverse plane, the color fields show decoherence at later times leading to the decay of the topological charge. 
The oscillating behavior of the axial charge density seen at earlier times is terminated by the decoherence.
We find that the decay of the fields rather helps nonzero axial charge persists to be present at later times.

The present work may provide an important basis for future investigations, which include first-principles-based simulations of the CME in real-time along the lines of \cite{Mueller:2016ven,Mace:2016shq}. It is also important to consider more realistic configurations of the gauge fields in heavy-ion collisions. In that case, the net axial charge would be vanishing after space or event averaging, and one has to investigate fluctuations of axial charge and their possible connections to observables. 

\acknowledgments

The author thanks Kenji Fukushima, Fran\c{c}ois Gelis, Larry McLerran and S\"{o}ren Schlichting for valuable discussions and comments. 
Part of numerical computation in this work was performed on the supercomputers JUQUEEN at J\"{u}lich Supercomputing Centre.


\end{document}